\pdfoutput=1

\documentclass[journal]{IEEEtran}
\usepackage{amsmath}
\usepackage{graphicx}
\usepackage{float}
\usepackage{cite}
\usepackage{color}
\usepackage{amssymb}
\usepackage{comment}    
\usepackage{subcaption}
\usepackage{enumitem}
\graphicspath{{pdfFigures/}}
\definecolor{lightpink}{rgb}    {1, 0.3, 0.9}
\definecolor{purple}{rgb}       {0.75, 0.45, 1}
\definecolor{lightblue}{rgb}    {0.3, 0.9, 1}
\definecolor{orange}{rgb}       {1, 0.65, 0.45}
\definecolor{lightgreen}{rgb}   {0.5, 1, 0.65}
\definecolor{yellow}{rgb}       {1, 0.8, 0}
\newcommand{\origin}{\boldsymbol{o}}

\begin{document}

\title{Sound Field Translation and Mixed Source Model for Virtual Applications with Perceptual Validation}

\author{
Lachlan Birnie\IEEEauthorrefmark{1}\IEEEauthorrefmark{4},
Thushara Abhayapala\IEEEauthorrefmark{1},
Vladimir Tourbabin\IEEEauthorrefmark{2},
Prasanga Samarasinghe\IEEEauthorrefmark{1}
    \thanks{\IEEEauthorrefmark{4}This research is supported by an Australian Government Research Training Program (RTP) Scholarship, and Facebook Reality Labs.}
\\
\IEEEauthorrefmark{1}Audio \& Acoustic Signal Processing Group, The Australian National University, Canberra, Australia
\\
\IEEEauthorrefmark{2}Facebook Reality Labs, Redmond, Washington, USA
}

\maketitle

\begin{abstract}
Non-interactive and linear experiences like cinema film offer high quality surround sound audio to enhance immersion, however the listener's experience is usually fixed to a single acoustic perspective. With the rise of virtual reality, there is a demand for recording and recreating real-world experiences in a way that allows for the user to interact and move within the reproduction. Conventional sound field translation techniques take a recording and expand it into an equivalent environment of virtual sources. However, the finite sampling of a commercial higher order microphone produces an acoustic sweet-spot in the virtual reproduction. As a result, the technique remains to restrict the listener's navigable region. In this paper, we propose a method for listener translation in an acoustic reproduction that incorporates a mixture of near-field and far-field sources in a sparsely expanded virtual environment. We perceptually validate the method through a Multiple Stimulus with Hidden Reference and Anchor (MUSHRA) experiment. Compared to the planewave benchmark, the proposed method offers both improved source localizability and robustness to spectral distortions at translated positions. A cross-examination with numerical simulations demonstrated that the sparse expansion relaxes the inherent sweet-spot constraint, leading to the improved localizability for sparse environments. Additionally, the proposed method is seen to better reproduce the intensity and binaural room impulse response spectra of near-field environments, further supporting the strong perceptual results. 
\end{abstract}


\section{Introduction}
\label{sec:introduction}

Virtual reality devices will provide a novel framework for people to interact with each other at a higher social bandwidth through immersive audio and visual reproductions of the real-world \cite{dodds2019auralization, suzuki20123d}. For example, in the future a person may be able to experience a live concert or orchestral performance through a virtual reproduction in their own home \cite{tylka2017models}. To complete the immersive experience, the listener/viewer should be allowed to explore and interact with the virtual reproduction \cite{amengual2019evaluation}. Subsequently, methods to accurately record and model the perceptual change in visual and auditory information as the user translates are required to maintain the original experience. 

Camera arrays have been used to capture visual information at multiple points-of-view for use in virtual reproductions \cite{zieglerimmersive}. Similarly, microphones distributed about an environment can record the spatial auditory scene from multiple points-of-view \cite{rivas2018practical}. However, hardware and feasibility restrictions limit the continuous space that can be recorded, and as a result, during reproduction the listener is usually stuck in the fixed acoustic perspective of the microphone \cite{salvador2017spatial}.

Recently, there have been two key approaches towards extending the auditory range that a listener can navigate inside a virtual sound field reproduction. These are an interpolation-based \cite{tylka2020fundamentals} and an extrapolation-based approach \cite{tylka2020performance}. Interpolation approaches utilize a grid of higher order microphones distributed about the acoustic space, and interpolate the sound to the listener during reproduction \cite{mariette2009sounddelta,tylka2019domains}. Better coloration and localization performance is expected from interpolation than extrapolation \cite{tylka2020fundamentals}. However, interpolation may not be feasible for all real-world scenarios due to the large spatial, hardware, and synchronization costs associated with constructing a microphone grid \cite{patricio2019toward}. Furthermore, typically listeners are confined to the interior region of the grid \cite{samarasinghe2014wavefield}, and sound sources within the grid may cause comb-filtering spectral distortions \cite{tylka2016soundfield}. Methods that alleviate these drawbacks and allow the listener to translate beyond the grid have been developed, however, they usually require additional localization and separation of direct sound field components \cite{wang2018translations,thiergart2013geometry}.

On the other hand, the extrapolation-based approach expands a single higher order microphone's recording outwards to the translated listener's position \cite{tylka2015comparison}. As a result, extrapolation overcomes many of the hardware and spatial drawbacks of the interpolation approach. Because a single microphone is utilized, the audio and visual capture system can occupy a single seat in the audience of a live event, which causes less obstruction and allows for more impromptu recordings.

Many extrapolation-based sound field translation methods have been developed to allow listener navigation in virtual reproduction, such as Ambisonic \cite{noisternig20033d,tylka2015comparison}, harmonic re-expansion \cite{menzies2007ambisonic}, discrete source \cite{pihlajamaki2015synthesis}, and point-source distribution \cite{fernandez2016sound}. One of the most popular extrapolation-based methods which we consider to be the benchmark in this paper, is the planewave method \cite{schultz2013data}. In this method, the recorded acoustic environment is expanded into a secondary distribution of virtual planewave sources \cite{duraiswami2005plane}. The secondary virtual environment constructs a sound field that is extrapolated from the microphone and is equivalent to the recording. The listener can perceptually move about the reproduction by translating the secondary environment's sound field \cite{schultz2013data,pihlajamaki2015synthesis}.

In practice, however, most extrapolation-based approaches, including the planewave method, are constrained by the higher order microphone used for recording \cite{tylka2015comparison}. Hardware limitations result in an approximate and truncated sound field recording that is confined to a finite region \cite{poletti2005three}, where the truncated recording is restricted by both the upper frequency band and the microphone radius \cite{ward2001reproduction}. As a result, the listener can only navigate within a small acoustic sweet-spot of a few centimeters which is defined by the commercial microphone's size \cite{acoustics2013em32}. Attempting to navigate beyond this inherent sweet-spot region, even after extrapolation, results in spectral distortions \cite{hahn2015modal,hahn2015physical,kuntz2007limitations}, degraded source localization \cite{winter2014localization,tylka2015comparison}, and a poor perceptual listening experience. 

Studies have shown that compensating for near-field effects attributes to better sound field reproduction \cite{daniel2003spatial}. Some reproduction methods have been able to model near-field sources with the use of prior knowledge of the source position or additional source localization processing \cite{wakayama2017extended,plinge2018six,thiergart2013geometry,kentgens2020translation}. However, the planewave translation method is limited by its far-field virtual source model, which makes the reproduction of near-field propagation difficult \cite{menzies2007nearfield}. 

In this paper we propose an alternative secondary source model for an extrapolated virtual reproduction that enables both a near-field and far-field propagation mixture. The method is built upon the benchmark planewave method which we review in Section \ref{sec:problem+planewave}. We expand the truncated recording of a commercial higher order microphone \cite{zyliaMic,visisonicsSphericalMic,acoustics2013em32} into a mixture of secondary virtual sources that are distributed in both the near-field and far-field (Section \ref{sec:MixedwaveMethod}), to create a more perceptually accurate reproduction. In addition to the source mixture, we also propose using a L1-norm regularization \cite{chartrand2008iteratively} to sparsely expand the recording into the equivalent virtual environment (Section \ref{sec:sparse}), as it has been shown to help extrapolate mode-limited sound fields \cite{emura2017sound,hu2019modeling,maeno2018mode}. 

We initially proposed the near-field far-field source mixture in \cite{birnie2019sound} without any substantial verification of the method's perceptual performance. In this paper, we study the perceptual aspects of the source mixture through a perceptual listening test with human subjects and investigate the results against numerical simulations of the extrapolated pressure and intensity fields. The perceptual evaluation presented in Section \ref{sec:PerceptionExperiment}, utilizes a MUltiple Stimulus with Hidden Reference and Anchor (MUSHRA) \cite{assemblyitu,ben2019loudness} framework adapted for use in a virtual environment to provide listeners with both an auditory and visual reference of the real-world environment \cite{patricio2019toward,plinge2018six}. We compare four translation methods with differing virtual source models and expansion techniques. We test the methods for the reproduction of human speech and music against the metrics of source localizability and robustness to spectral distortions. We will show that the proposed method offers greater perceptual accuracy and a more immersive experience for listeners moving throughout an expanded virtual reproduction. In Section \ref{sec:SimulationAnalysis}, we conduct a simulation study and show that the proposed method's perceptual performance is likely due to its ability to better reconstruct the near-field pressure and intensity of the original environment. We give our concluding remarks and suggestions for future work in Section \ref{sec:conclusion}. 


\section{Problem Formulation and the Planewave Sound Field Translation Method}
\label{sec:problem+planewave}

In this section, we formulate the problem of reconstructing a recorded real-world experience such that a listener is able to perceptually move through the acoustic reproduction. We first present the process of recording a general sound field with a commercial higher order microphone. We then review the planewave sound field translation method presented in \cite{schultz2013data}, which segments the reproduction into three parts. First, a virtual acoustic environment is built from a superposition of planewave sources. Second, planewave driving signals are estimated from the recording to model an equivalent acoustic environment. Third, a listener is placed inside the virtual equivalent environment and binaural signals are rendered as they move. We provide a discussion on the perceptual shortcomings of this planewave translation method at the end. 

\subsection{Sound Field Capture}

Consider a real-world acoustic environment that contains multiple sound sources, for example, a musical performance with many instruments. Let the origin $\origin$ denote the center of the environment's listening space, such as a seat in the middle of the audience. Each sound source is positioned at $\boldsymbol{z} = (r,\theta,\phi)$ with respect to $\origin$, where $\theta \in [0,\pi]$ is the elevation angle downwards from the z-axis, and $\phi \in [0,2\pi)$ is the azimuth angle counterclockwise from the x-axis. For a listener in the audience at position $\boldsymbol{d}$, the true sound they experience in the real-world can be described by 
\begin{equation}
\label{eq:reference}
    {}^\text{(real)}_\text{\{l,r\}}P(k,\boldsymbol{d}) 
    = 
    \sum_{\mu=1}^{U}
    H_\text{\{l,r\}}(k,\boldsymbol{z}_\mu;\boldsymbol{d}) 
    \times 
    s_\mu(k),
\end{equation}
where ${}^\text{(real)}_\text{\{l,r\}}P(k,\boldsymbol{d})$ is the pressure at the listener's left and right ear, $H_\text{\{l,r\}}(k,\boldsymbol{z};\boldsymbol{d})$ is the transfer function between each source and the listener's ears, or simply the Head-Related Transfer Function (HRTF) when the listener is in a free-field space without any reflections, $s_\mu(k)$ is the sound signal of the $\mu^\text{th}$ source, $\mu = (1,\cdots,U)$, $k = 2\pi f/c$ is the wave number, $f$ is the frequency, and $c$ is the speed of sound. From here on, we assume $H$ to be the free-field HRTF for simplicity. 

The aim is to record and reproduce the real-world auditory experience of (\ref{eq:reference}) for every possible listening position. The homogeneous sound field that encompasses every arbitrary listening position $\boldsymbol{x}$, where $|\boldsymbol{x}| < |\boldsymbol{z}|$, can be expressed through a spherical harmonic decomposition of \cite{eWilliams1999}
\begin{equation}
\label{eq:Psph}
    P(k,\boldsymbol{x}) = \sum_{n=0}^{\infty} \sum_{m=-n}^{n} \alpha_{nm}(k) j_n(k|\boldsymbol{x}|) Y_{nm}(\hat{\boldsymbol{x}}),
\end{equation}
where $|\cdot| \equiv r$, $\hat{\cdot} \equiv (\theta,\phi)$, $n$ and $m$ are index terms denoting spherical harmonic order and mode, respectively, $j_{n}(\cdot)$ are the spherical Bessel functions of the first kind, $Y_{nm}(\cdot)$ are the set of spherical harmonic basis functions, and $\alpha_{nm}(k)$ are the sound field's coefficients which completely describe the source-free acoustic environment centered about $\origin$ when $\alpha_{nm}(k)$ is known for all $n \in [0,\infty)$. 

In practice, the real-world acoustic environment can be recorded with an $N^\text{th}$ order microphone, by estimating the sound field's $\alpha_{nm}(k)$ coefficients for a finite set of $n \in [0,N]$. Consider a $N^\text{th}$ order microphone centered at $\origin$, such as a spherical \cite{acoustics2013em32} (or planar \cite{chen2015theory,visisonicsPlanarMic}) microphone array. The microphone array consists of $q = (1,\cdots,Q)$ pressure sensors that enclose the spherical acoustic region (listening space) of radius $|\boldsymbol{x}_Q|$ to be recorded. The sound field within this region can be estimated with \cite{abhayapala2002theory}
\begin{align}
\label{eq:alphaFromMic}
    & \alpha_{nm}(k) \approx \sum_{q=1}^{Q} w_q \frac{P(k,\boldsymbol{x}_q) Y_{nm}^*(\hat{\boldsymbol{x}}_q)}{b_{n}(k|\boldsymbol{x}_Q|)}, & n \in [0,N],
\end{align}
where $w_q$ are a set of suitable sampling weights \cite{rafaely2005analysis}, and $b_{n}(\cdot)$ is the rigid baffle equation \cite{eWilliams1999}. 

However, commercial $N^\text{th}$ order microphones can only record a small acoustic region $(|\boldsymbol{x}_Q| < 0.05\text{ m}$ \cite{acoustics2013em32}$)$ due to the hardware complexity and size constraint trade-offs with the spatial sampling Nyquist theorem \cite{poletti2005three}. The microphone's truncation order is restricted by the limited number of sensors, such that $Q \geq (N+1)^2$. Furthermore, the microphone's recording region and frequency range are balanced by the $N = \lceil k |\boldsymbol{x}_Q| \rceil$ rule \cite{kennedy2007intrinsic}. These two microphone properties define a maximum $|\boldsymbol{x}_Q|$ inside which the sound field is effectively of order $\leq N$. Beyond $|\boldsymbol{x}_Q|$, the reconstructed sound field requires higher orders $>N$ which are unknown, resulting in truncation error that degrades perceptual accuracy.

When reconstructing (\ref{eq:reference}) from the recording, the left and right ear signals for the listener can be reassembled in the spherical harmonic domain by \cite{zotkin2009regularized,zhang2010insights}
\begin{equation}
\label{eq:anchor}
    _\text{\{l,r\}}^\text{(mic)}P(k,\origin) = \sum_{n=0}^{N} \sum_{m=-n}^n H_\text{\{l,r\}}^{nm}(k) \times \alpha_{nm}(k),
\end{equation}
where $H_\text{\{l,r\}}^{nm}(k)$ are the spherical harmonic decomposition coefficients of the HRTF $H_\text{\{l,r\}}(k,\boldsymbol{z};\origin)$. In the reproduction (\ref{eq:anchor}), truncation forces the listener to the fixed auditory perspective of the microphone at $\origin$. If the listener attempts to move then they would immediately translate beyond the $|\boldsymbol{x}_Q|$ boundary and begin to experience spectral distortions, degraded source localization performance, and a loss in perceptual immersion. 

The objective of this paper is to relax this sweet-spot spatial constraint when reconstructing the sound field of a commercial microphone recording, and to build an equivalent virtual environment that allows for a listener to move about the acoustic space with a sustained perceptual immersion. For the remainder of this section we review the planewave sound field translation method that we consider to be the baseline method for enabling listener navigation.

\subsection{Planewave Distribution}
\label{sec:3.1}

The planewave sound field translation method aims to construct a virtual acoustic environment that is perceptually equivalent to the real-world recording. The building block of this virtual environment is the planewave source, whose sound field is modeled as
\begin{equation}
    P(k,\boldsymbol{x}) = \frac{e^{-ik\hat{\boldsymbol{y}}\cdot\boldsymbol{x}}}{4\pi},
\end{equation}
where $\hat{\boldsymbol{y}}$ denotes the planewave's incident direction. It is known that any acoustic free field can be modeled by an infinite superposition of planewaves \cite{duraiswami2005plane}. Therefore, the equivalent virtual environment is constructed from a spherical distribution of virtual planewave sources, expressed as
\begin{equation}
\label{eq:pwDist}
    ^\text{(pw)}P(k,\boldsymbol{x}) = \int \psi(k,\hat{\boldsymbol{y}};\origin) \frac{e^{-ik\hat{\boldsymbol{y}}\cdot\boldsymbol{x}}}{4\pi} d\hat{\boldsymbol{y}},
\end{equation}
where $\psi(k,\hat{\boldsymbol{y}};\origin)$ denotes the driving function of the planewave distribution as observed at $\origin$. If the driving function is modeled correctly then the planewave distribution can re-create the acoustic environment, such that ${}^\text{(pw)}P(k,\boldsymbol{x}) = {}^\text{(real)}P(k,\boldsymbol{x})$. To achieve this, the driving function needs to be estimated/expanded from the recorded $\alpha_{nm}(k)$ coefficients, which we describe next.

\subsection{Planewave Expansion}
\label{sec:3.2}

The sound field about $\origin$ due to a single virtual planewave can be expressed by the decomposition of \cite{eWilliams1999}
\begin{equation}
\label{eq:pwDecomp}
    \frac{e^{-ik\hat{\boldsymbol{y}}\cdot\boldsymbol{x}}}{4\pi} = \sum_{n=0}^{\infty} \sum_{m=-n}^{n} (-i)^n Y_{nm}^*(\hat{\boldsymbol{y}}) j_n(k|\boldsymbol{x}|) Y_{nm}(\hat{\boldsymbol{x}}).
\end{equation}
Additionally, the driving function centered at $\origin$ can also be expressed in terms of a harmonic decomposition, given as
\begin{equation}
\label{eq:psi=beta}
    \psi(k,\hat{\boldsymbol{y}};\origin) = \sum_{n'=0}^{\infty} \sum_{m'=-n'}^{n'} \beta_{n'm'}(k) Y_{n'm'}(\hat{\boldsymbol{y}}), 
\end{equation}
where $\beta_{nm}(k)$ are the spherical harmonic decomposition coefficients of $\psi(k,\hat{\boldsymbol{y}};\origin)$ which describe the sound field about the planewave distribution.
Substituting both (\ref{eq:pwDecomp}) and (\ref{eq:psi=beta}) into (\ref{eq:pwDist}) gives the planewave distribution's sound field in spherical harmonics, as
\begin{equation}
\label{eq:pw=beta}
    {}^\text{(pw)}P(k,\boldsymbol{x}) = 
    \sum_{n=0}^{\infty} \sum_{m=-n}^{n} 
    \underbrace{(-i)^n\beta_{nm}(k)}_{\alpha_{nm}(k)} 
    j_n(k|\boldsymbol{x}|) 
    Y_{nm}(\hat{\boldsymbol{x}}).
\end{equation}
From (\ref{eq:pw=beta}), the relationship between the $\beta_{nm}(k)$ coefficients and the recorded $\alpha_{nm}(k)$ coefficients can be extracted. Rearranging this relationship for $\beta_{nm}(k) = i^n\alpha_{nm}(k)$, expresses a planewave distribution that is equivalent to the recorded environment. Substituting this relationship back into (\ref{eq:psi=beta}), gives a closed-form expansion for a planewave driving function that matches the recording,
\begin{equation}
\label{eq:pwExp}
    \psi(k,\hat{\boldsymbol{y}};\origin) = \sum_{n=0}^{N} \sum_{m=-n}^{n} i^n \alpha_{nm}(k) Y_{nm}(\hat{\boldsymbol{y}}).
\end{equation}
Synthesizing a virtual environment with this driving function through (\ref{eq:pwDist}) produces a sound field that is equivalent to the recording. However, the recording (\ref{eq:alphaFromMic}) is only an approximation of the real environment, and therefore (\ref{eq:pwExp}) is also an approximate, such that  $^\text{(pw)}P(k,\boldsymbol{x}) \equiv {}^\text{(mic)}P(k,\boldsymbol{x}) \approx {}^\text{(real)}P(k,\boldsymbol{x})$.

\subsection{Planewave Auralization}
\label{sec:3.3}

A listener inside the planewave distribution is immersed within a spatial reproduction of the real-world acoustic environment. The binaural signals for the listener at the distribution center can be presented by exchanging their HRTF into (\ref{eq:pwDist}), giving
\begin{equation}
\label{eq:pwLR1}
    ^\text{(pw)}_\text{\{l,r\}}P(k,\origin) = \int \psi(k,\hat{\boldsymbol{y}};\origin) H_\text{\{l,r\}}(k,\hat{\boldsymbol{y}};\origin) d\hat{\boldsymbol{y}}.
\end{equation}
Furthermore, the planewave distribution allows for the listener to move perceptually about the reproduction. 
The sound heard by the listener who is translated to $\boldsymbol{x} = [\origin + \boldsymbol{d}] \equiv \boldsymbol{d}$ can be derived from (\ref{eq:pwDist}) as
\begin{equation}
\begin{split}
\label{eq:pwTransDist}
    ^\text{(pw)}P(k,[\origin + \boldsymbol{d}]) 
    & = \int \psi(k,\hat{\boldsymbol{y}};\origin) \frac{e^{-ik\hat{\boldsymbol{y}}\cdot[\origin+\boldsymbol{d}]}}{4\pi} d\hat{\boldsymbol{y}} \\
    & = \int \psi(k,\hat{\boldsymbol{y}};\origin) e^{-ik\hat{\boldsymbol{y}}\cdot\boldsymbol{d}} \frac{e^{-ik\hat{\boldsymbol{y}}\cdot\origin}}{4\pi} d\hat{\boldsymbol{y}}.
\end{split}
\end{equation}
It is observed from (\ref{eq:pwTransDist}) that the translation in space differs only by a phase shift in the planewave driving function. Therefore, applying the translational phase shift of \cite{menzies2007nearfield}
\begin{equation}
\label{eq:phaseShift}
    \psi(k,\hat{\boldsymbol{y}};\boldsymbol{d}) = \psi(k,\hat{\boldsymbol{y}};\origin) \times e^{-ik\hat{\boldsymbol{y}}\cdot\boldsymbol{d}},
\end{equation}
to the binaural signals in (\ref{eq:pwLR1}), allows for the listener to dynamically move their acoustic perspective by
\begin{equation}
\label{eq:pwAural}
    ^\text{(pw)}_\text{\{l,r\}}P(k,\boldsymbol{d}) 
    = \int \psi(k,\hat{\boldsymbol{y}};\boldsymbol{d})  H_\text{\{l,r\}}(k,\hat{\boldsymbol{y}};\origin) d\hat{\boldsymbol{y}}.
\end{equation}

In practice, the virtual planewave distribution (\ref{eq:pwDist}) can be realized with a discrete set of known sources, 
\begin{equation}
    ^\text{(pw)}P(k,\boldsymbol{x}) \approx \sum_{\ell=1}^{L} w_\ell \psi(k,\hat{\boldsymbol{y}}_\ell;\origin) \frac{e^{-ik\hat{\boldsymbol{y}}_\ell\cdot\boldsymbol{x}}}{4\pi}
\end{equation}
where $\ell = (1,\cdots,L)$ index each virtual planewave, $L$ is the total number of sources, and $w_\ell$ are a set of suitable sampling weights. Similarly, the dynamic binaural signals can be realized from the discrete distribution with 
\begin{equation}
\label{eq:pwMethod}
    ^\text{(pw)}_\text{\{l,r\}}P(k,\boldsymbol{d}) 
    \approx \sum_{\ell=1}^{L} w_\ell \psi(k,\hat{\boldsymbol{y}}_\ell;\boldsymbol{d})  H_\text{\{l,r\}}(k,\hat{\boldsymbol{y}}_\ell;\origin).
\end{equation}

\begin{figure}
    \centering
    \includegraphics[width=\columnwidth]{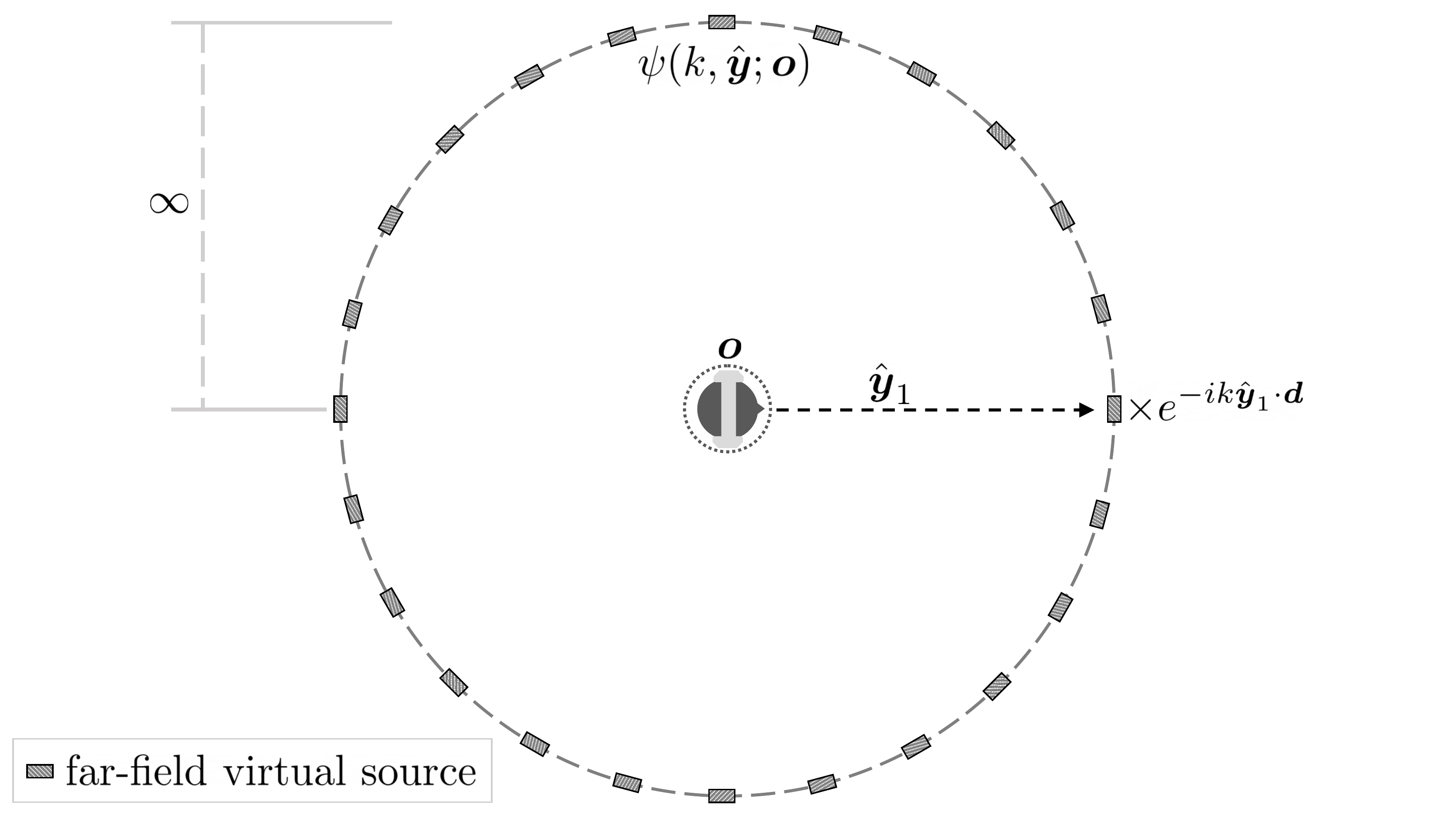}
    \caption{Illustration of the equivalent virtual planewave distribution. The listener's perspective is fixed at the distribution center $\origin$, where a phase shift applied to the driving function translates the sound field about the listener.}
    \label{fig:pwMethod}
\end{figure}

We illustrate this planewave method to sound field translation in Fig. \ref{fig:pwMethod}. The reproduction is expressed by many discrete planewave signals that are known continuously throughout the virtual environment. Therefore, the planewave method does not explicitly limit the amount the listener can translate. However, (\ref{eq:pwMethod}) uses $\psi(k,\hat{\boldsymbol{y}};\origin)$ which is estimated through (\ref{eq:alphaFromMic}) and (\ref{eq:pwExp}). As a result, the $N^\text{th}$ order truncation inherently remains, and the listener's movement is still implicitly limited. 

\subsection{Discussion}
\label{sec:pwDiscussion}
The virtual planewave expansion enables listener translation, however, some shortcomings are still exhibited in the listener's perception: 
\begin{itemize}
    \item As mentioned, the planewave method inherits truncation artifacts through an over-approximation of (\ref{eq:pwExp}), and the listener's movement remains inherently restricted inside the virtual reproduction. As the listener translates further away from the recording's sweet-spot, they begin to experience spectral distortions, a loss in source localization, and poorer perceptual accuracy. 
    
    \item The planewave expansion has difficulties in synthesizing near-field sound sources due to its far-field source model. 

    \item The planewave auralization (\ref{eq:pwMethod}) fixes the HRTF perspective to the virtual distribution center, $H_\text{\{l,r\}}(k,\hat{\boldsymbol{y}}_\ell;\origin)$, and performs translation by phase shifting the sound field with (\ref{eq:phaseShift}). However, the HRTF propagation vectors $\hat{\boldsymbol{y}}_\ell;\origin$ remain un-shifted, and as a result, the HRTF models un-translated head reflections as the listener moves.  
    
\end{itemize}
In the next section we propose an alternative sound field translation model to address the above shortcomings.

\section{Mixedwave Sound Field Translation Method}
\label{sec:MixedwaveMethod}

In this section, we define a virtual source that models both a near-field and far-field propagation, which we will refer to as a mixedwave source. We then build a virtual distribution of mixedwave sources and expand a real-world recording into an equivalent sound field. Additionally, we also propose a sparse method for expanding a virtual source distribution that alleviates some of the spatial restrictions imposed by the truncated recording. 

\subsection{Mixture of Near-Field and Far-Field Sources}

Here, we define the virtual source that will be the building block for our proposed method. Consider a near-field point-source at $\boldsymbol{y}$, where the driving signal of the source with respect to itself is denoted $\Dot{\psi}(k,\boldsymbol{y})$. We can express the driving function observed at a position $\boldsymbol{x}$ with \cite{eWilliams1999}
\begin{equation}
\label{eq:psDist1}
    \psi(k,\boldsymbol{y};\boldsymbol{x}) = \Dot{\psi}(k,\boldsymbol{y}) \frac{e^{ik||\boldsymbol{y} - \boldsymbol{x}||}}{||\boldsymbol{y} - \boldsymbol{x}||}.
\end{equation}
Evaluating (\ref{eq:psDist1}) when $\boldsymbol{x} = \origin$ gives the driving function observed by a receiver/microphone, as
\begin{equation}
\label{eq:psDist2}
    \psi(k,\boldsymbol{y};\origin) = \Dot{\psi}(k,\boldsymbol{y}) \frac{e^{ik|\boldsymbol{y}|}}{|\boldsymbol{y}|}.
\end{equation}
Rearranging (\ref{eq:psDist2}) gives an expression for the source signal in terms of the source's distance and the driving function observed by the receiver,
\begin{equation}
\label{eq:psDist3}
    \Dot{\psi}(k,\boldsymbol{y}) = \psi(k,\boldsymbol{y};\origin) |\boldsymbol{y}| e^{-ik|\boldsymbol{y}|}.
\end{equation}
Substituting (\ref{eq:psDist3}) back into (\ref{eq:psDist1}) provides the driving function observed at any arbitrary point $\boldsymbol{x}$ in terms of the function observed by the receiver/microphone, expressed as
\begin{equation}
    \psi(k,\boldsymbol{y};\boldsymbol{x}) = \underbrace{\psi(k,\boldsymbol{y};\origin) |\boldsymbol{y}| e^{-ik|\boldsymbol{y}|}}_{\Dot{\psi}(k,\boldsymbol{y})} \frac{e^{ik||\boldsymbol{y} - \boldsymbol{x}||}}{||\boldsymbol{y} - \boldsymbol{x}||}.
\end{equation}
We note that the $|\boldsymbol{y}|e^{-ik|\boldsymbol{y}|}$ term can be seen to have redefined the point-source from being a function with respect to itself, to being a function with respect to $\origin$. This allows us to observe the source distribution at $\origin$ with a microphone and estimate the sound at any translated position $\boldsymbol{x}$.

Additionally, the constant term has the property of \cite{ward2001reproduction}
\begin{equation}
\label{eq:pointLimit}
    \lim_{|\boldsymbol{y}| \rightarrow \infty} |\boldsymbol{y}|e^{-ik|\boldsymbol{y}|} \frac{e^{ik||\boldsymbol{y}-\boldsymbol{x}||}}{4\pi||\boldsymbol{y} - \boldsymbol{x}||} = \frac{e^{-ik\hat{\boldsymbol{y}}\cdot\boldsymbol{x}}}{4\pi},
\end{equation}
which allows for a mixture of near-field and far-field virtual source distributions to be modeled with this building block. We define this building block as the mixedwave source, 
\begin{equation}
\label{eq:mwSource}
    P(k,\boldsymbol{x}) = |\boldsymbol{y}| e^{-ik|\boldsymbol{y}|} \frac{e^{ik||\boldsymbol{y}-\boldsymbol{x}||}}{4\pi||\boldsymbol{y}-\boldsymbol{x}||}.
\end{equation}
In the spherical harmonic domain,
\begin{equation}
\label{eq:mwDecomp}
\begin{split}
    & |\boldsymbol{y}| e^{-ik|\boldsymbol{y}|} \frac{e^{ik||\boldsymbol{y}-\boldsymbol{x}||}}{4\pi||\boldsymbol{y}-\boldsymbol{x}||}
    = \\ &
    \sum_{n=0}^{\infty} \sum_{m=-n}^{n}  ik|\boldsymbol{y}|e^{-ik|\boldsymbol{y}|} h_n(k|\boldsymbol{y}|) Y_{nm}^*(\hat{\boldsymbol{y}})
    j_n(k|\boldsymbol{x}|) Y_{nm}(\hat{\boldsymbol{x}}),
\end{split}
\end{equation}
where $h_n(\cdot)$ is the spherical Hankel function of the first kind. We note that spherical Hankel functions also have
\begin{equation}
\label{eq:mwLimSph}
    \lim_{|\boldsymbol{y}| \rightarrow \infty} ik |\boldsymbol{y}|e^{-ik|\boldsymbol{y}|} h_n(k|\boldsymbol{y}|) = (-i)^n,
\end{equation}
to correspond with (\ref{eq:pointLimit}). We can observe from (\ref{eq:mwLimSph}) that when the mixedwave source is placed in the far-field, the definition of (\ref{eq:mwDecomp}) will match that of the planewave source (\ref{eq:pwDecomp}). This property then allows for both a near-field sound propagation to be modeled by a mixedwave distribution with a small radius, and a far-field sound propagation modeled by a mixedwave distribution with a large radius. We will use this near-field and far-field distribution of mixedwave sources as the basis of our proposed sound field translation method next. 

\subsection{Mixedwave Method for Sound Field Translation}

Following the planewave translation method, our proposed mixedwave translation method is also broken into three parts. 

\subsubsection{Mixedwave Distribution}

We propose constructing a virtual equivalent sound field from two concentric spherical distributions of mixedwave sources. The first virtual sphere is placed in the near-field with a radius of $R_\text{(nf)}$, and the second sphere is placed at $R_\text{(ff)}$ in the far-field, such that 
\begin{equation}
\label{eq:mwDist}
\begin{split}
    & ^\text{(mw)}P(k,\boldsymbol{x})
    \\ & = \int \psi(k,R_\text{(nf)}\hat{\boldsymbol{y}};\origin) R_\text{(nf)} e^{-ikR_\text{(nf)}} 
    \frac{e^{ik||R_\text{(nf)}\hat{\boldsymbol{y}}-\boldsymbol{x}||}}
    {4\pi||R_\text{(nf)}\hat{\boldsymbol{y}}-\boldsymbol{x}||} d\hat{\boldsymbol{y}}
    \\ & + \int \psi(k,R_\text{(ff)}\hat{\boldsymbol{y}};\origin) R_\text{(ff)} e^{-ikR_\text{(ff)}}
    \frac{e^{ik||R_\text{(ff)}\hat{\boldsymbol{y}}-\boldsymbol{x}||}}
    {4\pi||R_\text{(ff)}\hat{\boldsymbol{y}}-\boldsymbol{x}||} d\hat{\boldsymbol{y}},
\end{split}
\end{equation}
where, $\psi(k,R\hat{\boldsymbol{y}};\origin)$, $R \in \{R_\text{(nf)}, R_\text{(ff)}\}$, are the driving functions of the two mixedwave distributions centered at $\origin$.

\subsubsection{Mixedwave Expansion}

Following the procedure in Section \ref{sec:3.2}, we can decompose the $\psi(k,R\hat{\boldsymbol{y}};\origin)$ driving function into spherical harmonic aperture coefficients of $\beta_{n'm'}(k,R)$, expressed as
\begin{equation}
\label{eq:psi=betaR}
    \psi(k,R\hat{\boldsymbol{y}};\origin) = \sum_{n'=0}^{\infty} \sum_{m'=-n'}^{n'} \beta_{n'm'}(k,R) Y_{n'm'}(\hat{\boldsymbol{y}}).
\end{equation}
We substitute both (\ref{eq:psi=betaR}) and (\ref{eq:mwDecomp}) into (\ref{eq:mwDist}) to extract the relationship between $\beta_{nm}(k)$ and $\alpha_{nm}(k)$, given as
\begin{equation}
\label{eq:beta=alpha2}
    \beta_{nm}(k,R) = \sum_{n=0}^{N} \sum_{m=-n}^{n} \frac{\alpha_{nm}(k)}{ikRe^{-ikR}h_n(kR)}.
\end{equation}
Finally, we substitute (\ref{eq:beta=alpha2}) back into (\ref{eq:psi=betaR}) to derive a closed-form expansion for the mixedwave driving functions in terms of the recorded coefficients,
\begin{equation}
\label{eq:mwExp}
    \psi(k,R\hat{\boldsymbol{y}};\origin) = \sum_{n=0}^{N} \sum_{m=-n}^{n} \frac{\alpha_{nm}(k)}{ikRe^{-ikR}h_n(kR)} Y_{nm}(\hat{\boldsymbol{y}}).
\end{equation}
We use a set of real-world recorded coefficients $\alpha_{nm}(k)$ with (\ref{eq:mwExp}) to estimate the driving functions of the near-field and far-field virtual distributions, such that ${}^\text{(mw)}P(k,\boldsymbol{x}) \equiv {}^\text{(mic)}P(k,\boldsymbol{x}) \approx {}^\text{(real)}P(k,\boldsymbol{x})$.

\subsubsection{Mixedwave Auralization}

\begin{figure}
    \centering
    \includegraphics[width=\columnwidth]{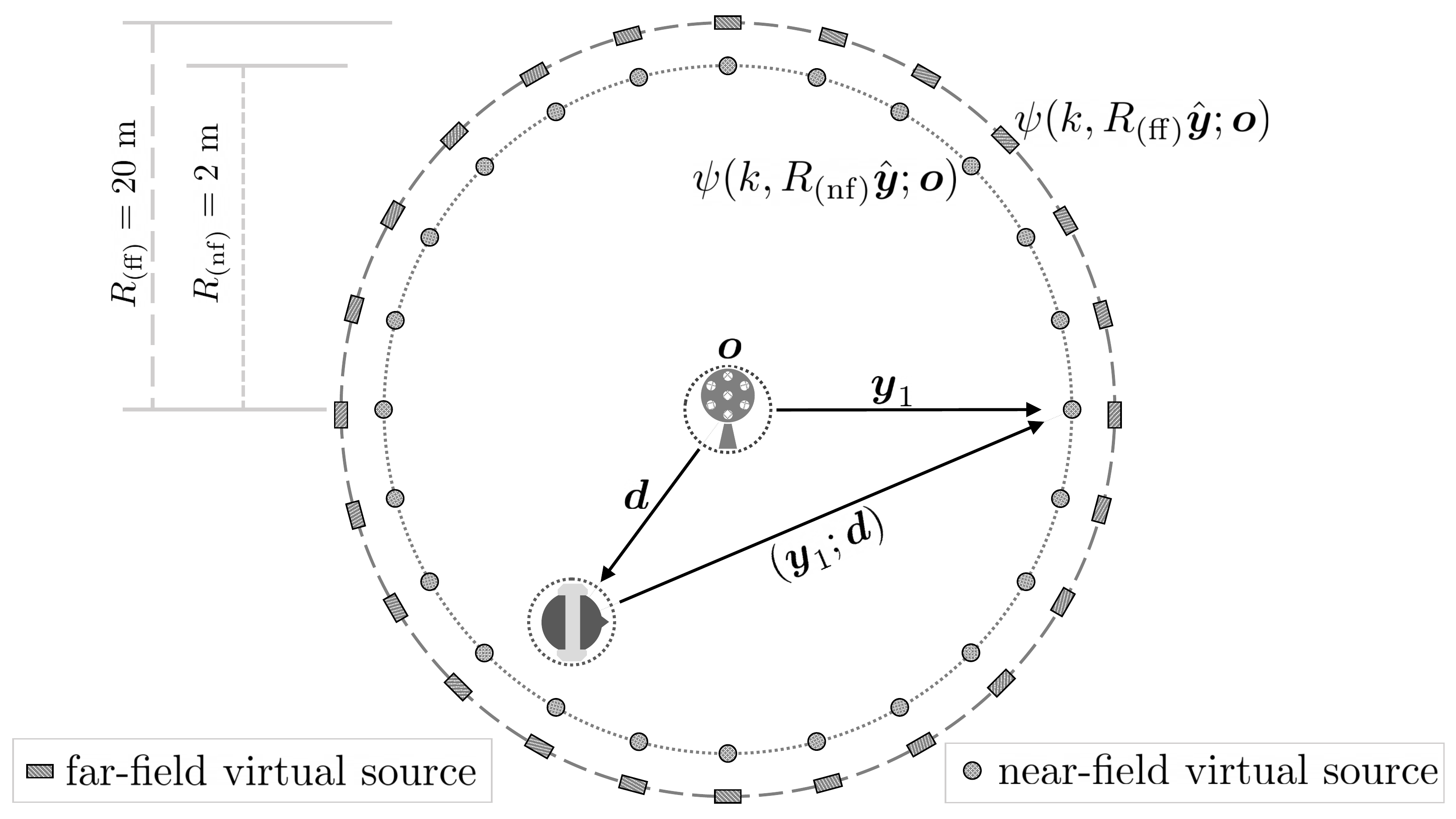}
    \caption{Illustration of the equivalent virtual mixedwave sound field. The listener is translated to $\boldsymbol{d}$, and the vectors $(\boldsymbol{y}_\ell;\boldsymbol{d})$ are updated with the HRTF to auralize an immersive reproduction.}
    \label{fig:mwMethod}
\end{figure}

Consider a listener inside the virtual mixedwave distribution at the translated position $\boldsymbol{x} = [\origin + \boldsymbol{d}] \equiv \boldsymbol{d}$, $|\boldsymbol{d}| < R_\text{(nf)}$, as shown in Fig. \ref{fig:mwMethod}. We render the left and right binaural signals by applying the mixedwave driving function to the HRTF based on the listener's translated position, given as \cite{tylka2015comparison}
\begin{equation}
    ^\text{(mw)}_\text{\{l,r\}}P(k,\boldsymbol{d}) = \int \psi(k,R\hat{\boldsymbol{y}};\origin) H_\text{\{l,r\}}(k,R\hat{\boldsymbol{y}};\boldsymbol{d}) d\boldsymbol{y},
\end{equation}
where $R\hat{\boldsymbol{y}};\boldsymbol{d}$ denotes the propagation direction of the mixedwave source with respect to $\boldsymbol{d}$, which is given by $(\boldsymbol{y} - \boldsymbol{d})$. We note that this is possible for the mixedwave distribution due to the finite positions of each source, unlike the infinite definitions of planewave sources. 

Once again, a set of discrete sources can be used to practically realize the virtual mixedwave distributions, expressed as $^\text{(mw)}P(k,\boldsymbol{x}) \approx$
\begin{equation}
\begin{split}
    & \sum_{\ell=1}^{L} 
    w_\ell 
    \psi(k,R_\text{(nf)}\hat{\boldsymbol{y}}_\ell;\origin) 
    R_\text{(nf)} e^{-ikR_\text{(nf)}}
    \frac{e^{ik||R_\text{(nf)}\hat{\boldsymbol{y}}_\ell-\boldsymbol{x}||}}{4\pi||R_\text{(nf)}\hat{\boldsymbol{y}}_\ell-\boldsymbol{x}||}
    \\ + 
    & \sum_{\ell=1}^{L} 
    w_\ell 
    \psi(k,R_\text{(ff)}\hat{\boldsymbol{y}}_\ell;\origin) 
    R_\text{(ff)} e^{-ikR_\text{(ff)}} 
    \frac{e^{ik||R_\text{(ff)}\hat{\boldsymbol{y}}_\ell-\boldsymbol{x}||}}{4\pi||R_\text{(ff)}\hat{\boldsymbol{y}}_\ell-\boldsymbol{x}||},
\end{split}
\end{equation}
 where the near-field and far-field distributions each contain $L$ sources. Similarly, we realize the mixedwave auralization within the discrete virtual distributions by
\begin{equation}
\label{eq:mwMethod}
    ^\text{(mw)}_\text{\{l,r\}}P(k,\boldsymbol{d}) = \sum_{\ell=1}^{2L} w_\ell \psi(k,\boldsymbol{y}_\ell;\origin) H_\text{\{l,r\}}(k,\boldsymbol{y}_\ell;\boldsymbol{d}),
\end{equation}
where $|\boldsymbol{y}_\ell|  = R_\text{(nf)}$ for $\ell \in [1,L]$, and $|\boldsymbol{y}_\ell|  = R_\text{(ff)}$ for $\ell \in [L+1,2L]$, and $\boldsymbol{y}_\ell;\boldsymbol{d}$ is the propagation direction of the $\ell^\text{th}$ mixedwave source with respect to the translated listener. Unlike the planewave method, the maximum distance a listener can translate within the mixedwave environment is restricted by $R_\text{(nf)}$. However, we suspect that $R_\text{(nf)}$ can be selected to match the size of a small real-world room that is recorded.

\subsection{Sparse Expansion Methods}
\label{sec:sparse}

The closed-form expansion constructs a virtual environment that is equivalent to the original recording. However. the expansion distributes energy $\psi(k,\boldsymbol{y}_\ell;\origin)$ throughout all virtual sources. This causes an over-approximation of the truncated recording's underlying spatial artifacts. As a result, the amount a listener can translate before experiencing a loss in immersion is still inherently restricted by the recording's truncation. Furthermore, it is believed that modeling fewer virtual sources from propagation directions that are similar to the original environment will lead to better perceptual immersion \cite{birnie2019sound}. For these reasons, we propose a sparse constrained expansion method for constructing our virtual mixedwave environment. 

The coefficients $\alpha_{nm}(k)$ observed at the center of a virtual distribution can be expressed in matrix form as
\begin{equation}
\label{eq:APsi=alpha}
    \boldsymbol{A}\boldsymbol{\psi} = \boldsymbol{\alpha},
\end{equation}
where $\boldsymbol{\alpha} = [\alpha_{00}(k), \alpha_{1-1}(k), \cdots, \alpha_{NN}(k)]^\text{T}$ are the recorded coefficients, $\boldsymbol{\psi} = [\psi(k,\boldsymbol{y}_{1};\origin), \cdots, \psi(k,\boldsymbol{y}_{\mathcal{L}};\origin)]$ are the $\mathcal{L}$ equivalent virtual source driving signals, and $\boldsymbol{A}$ is the $(N+1)^2 \text{ by } \mathcal{L}$ expansion matrix. The entries of $\boldsymbol{A}$ are given by $(-i)^n Y_{nm}^{*}(\hat{\boldsymbol{y}}_\ell)$ for a planewave expansion (\ref{eq:pwExp}), and $ik|\boldsymbol{y}_\ell|e^{-ik|\boldsymbol{y}_\ell|} h_n(k|\boldsymbol{y}_\ell|) Y_{nm}^{*}(\hat{\boldsymbol{y}}_\ell)$ where $\mathcal{L} = 2L$ for the two source distributions of a mixedwave expansion (\ref{eq:mwExp}). We assume $L > (N+1)^2$ for the under-determined case. 

We construct a sparse source distribution by solving the linear regression problem (\ref{eq:APsi=alpha}) using Iteratively Reweighted Least Squares (IRLS) \cite{chartrand2008iteratively}. In brief, the IRLS approach replaces the $\ell^\text{p}$-objective function
\begin{equation}
    \min_{\boldsymbol{\psi}} ||\boldsymbol{\psi}||_\text{p}^\text{p} \hspace{0.5cm} \text{ subject to } \boldsymbol{A}\boldsymbol{\psi} = \boldsymbol{\alpha},
\end{equation}
with a weighted $\ell^2$-norm,
\begin{equation}
\label{eq:IRLS1}
    \min_{\boldsymbol{\psi}} \sum_{i=1}^{\mathcal{L}} w_i\boldsymbol{\psi}_{i}^{2}
    \hspace{0.5cm} \text{ subject to } \boldsymbol{A}\boldsymbol{\psi} = \boldsymbol{\alpha},
\end{equation}
where $w_{i} = |\boldsymbol{\psi}_{i}^{(\nu-1)}|^{p-2}$ are the weights computed from the previous iterate $\boldsymbol{\psi}^{(\nu-1)}$. The next iterate is given by
\begin{equation}
    \boldsymbol{\psi}^{(\nu)} = Q_\nu \boldsymbol{A}^\text{T} \left( \boldsymbol{A} Q_{\nu} \boldsymbol{A}^\text{T} \right)^{-1} \boldsymbol{\alpha},
\end{equation}
where $Q_{\nu}$ is the diagonal matrix with $1/w_{i} = |\boldsymbol{\psi}_{i}^{(\nu-1)}|^{2-p}$. Other regularization techniques can also be utilized, such as the Least-Absolute Shrinkage and Selection Operator (Lasso) \cite{lilis2010sound,tibshirani1996regression}, and we direct the reader to \cite{candes2008introduction} for further information in regards to compressive sensing. 

\subsection{Discussion} 
Continuing our discussion on the planewave method's shortcoming in Sec. \ref{sec:pwDiscussion}, we give the following comments:
\begin{itemize}
    \item Sparsely expanding the virtual source distribution (\ref{eq:APsi=alpha}) with IRLS is expected to further enhance the perceptual immersion for a listener, as they should experience more localized virtual sources. Additionally, the sparsity relaxes the spatial sweet-spot restriction and over-approximation issue stemming from the closed-form expansion used by the planewave method. These properties are demonstrated by experiment in Section \ref{sec:PerceptionExperiment} and by simulation in Section \ref{sec:SimulationAnalysis}. 

    \item The mixedwave distribution can easily synthesize near-field sound sources. The modified point-source (\ref{eq:mwSource}) can model a spherical-wave propagation by simply positioning the mixedwave source in the near-field. 
    
    \item The mixedwave auralization (\ref{eq:mwMethod}) translates the HRTF with the listener. As a result, the propagation vectors in  $H_\text{\{l,r\}}(k,\boldsymbol{y}_\ell;\boldsymbol{d})$ are updated with $\boldsymbol{d}$ to render changes in head reflection, similar to virtual higher-order Ambisonics \cite{tylka2015comparison}. 
    Intuitively, this is expected to result in greater perceptual immersion. 
\end{itemize}
We examine the perceptual advantages of the sparse expansion and the mixedwave source model against the planewave benchmark experimentally in the next section.


\section{Perceptual Experiment} 
\label{sec:PerceptionExperiment}
\newcommand{\xyHeight}{1.25\text{ m}}
\newcommand{\micR}{0.042\text{ m}}
\newcommand{\srcPos}{(1, 0, 0)\text{ m}}
\newcommand{\micPos}{(0, 0, 0)\text{ m}}

Our aim is to maintain the immersion for a listener inside an acoustic reproduction. Therefore, it is of crucial importance, foremost, that we evaluate the proposed method against the planewave benchmark in a perceptual listening experiment. This section outlines the perceptual experiment system we implemented and presents the statistical results at the end.

\subsection{Experiment Methodology}

\subsubsection{Compared Methods}

We conducted a MUSHRA perceptual experiment to compare four translation methods. In total the experiment presented six signals:
\begin{itemize}
    \item \emph{Reference / hidden reference:} Signals of the true free-field transfer function between a real-world point-source and the translated listener, given by (\ref{eq:reference}).
    
    \item \emph{Anchor:} Signals of the truncated recording that is fixed spatially to the microphone's position (\ref{eq:anchor}). Sound field rotation is still rendered, but no translation is processed. This is a similar anchor to the three-degrees-of-freedom used in \cite{plinge2018six}. 
    
    \item \emph{Benchmark / planewave closed-form (PW-CF):} Signals rendered of a virtual planewave distribution (\ref{eq:pwMethod}) that are expanded through the closed-form expression (\ref{eq:pwExp}).
    
    \item \emph{Planewave IRLS (PW-IRLS):} Signals of a IRLS (Section \ref{sec:sparse}) sparsely expanded planewave distribution.
    
    \item \emph{Mixedwave closed-form (MW-CF):} Signals rendered of a virtual mixedwave distribution (\ref{eq:mwMethod}) that are expanded through the closed-form expression (\ref{eq:mwExp}). 
    
    \item \emph{Proposed method / mixedwave IRLS (MW-IRLS):} Signals of a IRLS sparsely expanded mixedwave distribution. 
\end{itemize}

The experiment comprised of four tests with two scoring metrics, \emph{source localization} and \emph{basic audio quality}, and two sound-signals, \emph{speech} and \emph{music}. The source localization test asked listeners to score on the perceived direction of the sound-source, the source width, and the sound field sparseness with respect to both a visual-reference and the reference signal. Whereas, the basic audio quality test asked listeners to score against the reference for spectral distortions and other audible processing artifacts. In total the scores of 17 participants were collected for the speech sound-source, and 11 scores for the music sound-source. The recording microphone was shown in the virtual environment, and listeners were informed that the further they translate, the greater the differences they should perceive between methods. We asked the listeners to score while accounting for each method's performance over a $1\text{ m}$ square reproduction space.

\subsubsection{Experiment System}

We used an Oculus Rift along with a pair of Beyerdynamic DT 770 pro headphones to track the listener and provide a visual reference of the true sound source. We used the HRTFs of the FABIAN head and torso simulator \cite{lindau2007binaural} from the HUTUBS dataset \cite{brinkmann2019cross,fabian2019hutubs} for auralization. The HRTFs were rotated for each test signal by multiplying the HRTF coefficients with Wigner-D functions \cite{rafaely2008spherical}. Signals were processed at a frame size of $4096$ with $50\%$ overlap and a $16\text{ kHz}$ sampling frequency, due to hardware constraints and the computational costs of the real-time experiment. 

\subsubsection{Virtual Environment}
\label{sec:virtualEnvironment}

We simulated the real-world auditory experience with a single free-field point-source in order to generate a true experiment reference signal for the listener at every position. We constructed a virtual environment with $\origin$ placed at the center, and the XY-plane $\xyHeight$ above the ground to align with a listener's head while sitting. We modeled the \emph{true} sound-source with a static point-source at $\srcPos$. By \emph{true}, we signify that the sound field generated by this point-source is denoted as the real-world auditory experience we record and reproduce. 

Additionally, we also simulated the process of recording the truncated sound field of the true point-source. We used a $4^\text{th}$ order rigid 36-sensor spherical microphone array centered at $\origin$. 
Microphone sensors were distributed at Fliege positions \cite{fliege1999distribution} with $\micR$ radius to best represent a commercial microphone \cite{acoustics2013em32}. 
Recordings were generated by convolving the sound-source's signal with the microphone's impulse response. The $\alpha_{nm}(k)$ coefficients were extracted with (\ref{eq:alphaFromMic}) before being expanded into virtual distributions. 

The planewave distribution consisted of $L = 36$ virtual sources at Fliege positions \cite{fliege1999distribution}. This selection was made as a trade-off with computation complexity. However, adding more planewaves is not expected to improve source localization performance, as the distribution already over-samples the $4^\text{th}$ order recording \cite{winter2014localization,hahn2015physical,tylka2020performance}. Similarly, the mixedwave distribution consisted of two sets of $L = 36$ virtual sources at the same Fliege positions. The first set was distributed in the near-field at $R_\text{(nf)} = 2\text{ m}$, and the second was placed at $R_\text{(ff)} = 20\text{ m}$ in the far-field.

\subsubsection{Experiment Auralization}

The reference was rendered by convolving (in frequency domain) the sound-source signal with the true source-to-listener HRTF. For the anchor (\ref{eq:anchor}), the signals were convolved by multiplying $\alpha_{nm}(k)$ with the spherical HRTF-coefficients directly \cite{zhang2010insights}. The planewave method signals were rendered with the convolution of the HRTF at $\origin$ and the phase-shifted driving function (\ref{eq:pwMethod}). The phase-shift was updated with the Oculus head position to render perceptual translation. For the mixedwave methods, the HRTFs were reconstructed between each source and the listener's translated position. Binaural signals were then rendered with the convolution of the mixedwave driving function and the $\boldsymbol{y}_\ell$-to-$\boldsymbol{d}$ HRTF (\ref{eq:mwMethod}).

\newcommand{\boxScale}{0.49}
\begin{figure*}
    \centering 
    \begin{subfigure}{\boxScale\textwidth}
        \centering
        \includegraphics[width=\columnwidth]{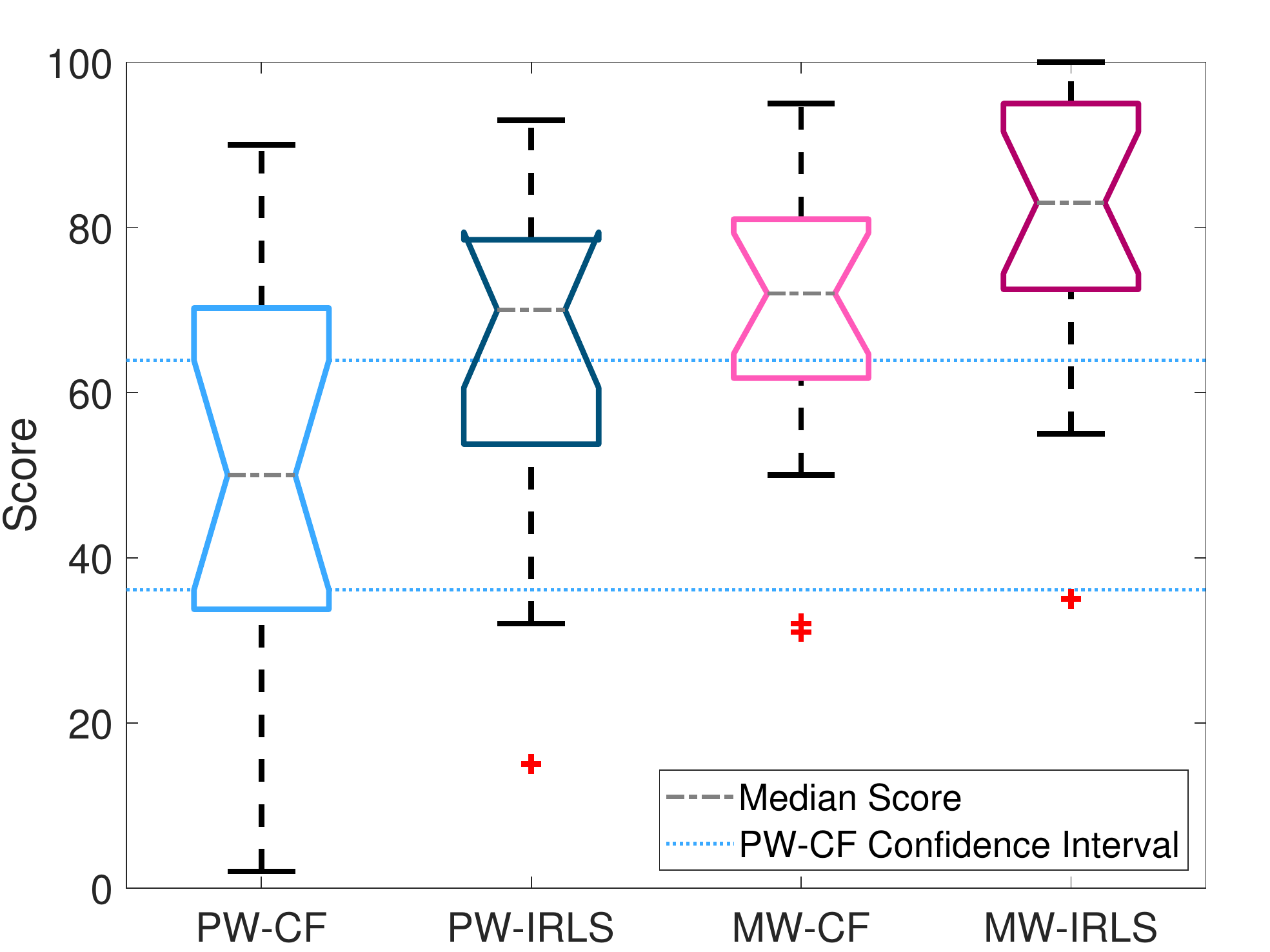}
        \caption{Source localization scores with speech sound-source.}
        \label{fig:spc-dir}
    \end{subfigure}
    \begin{subfigure}{\boxScale\textwidth}
        \centering
        \includegraphics[width=\columnwidth]{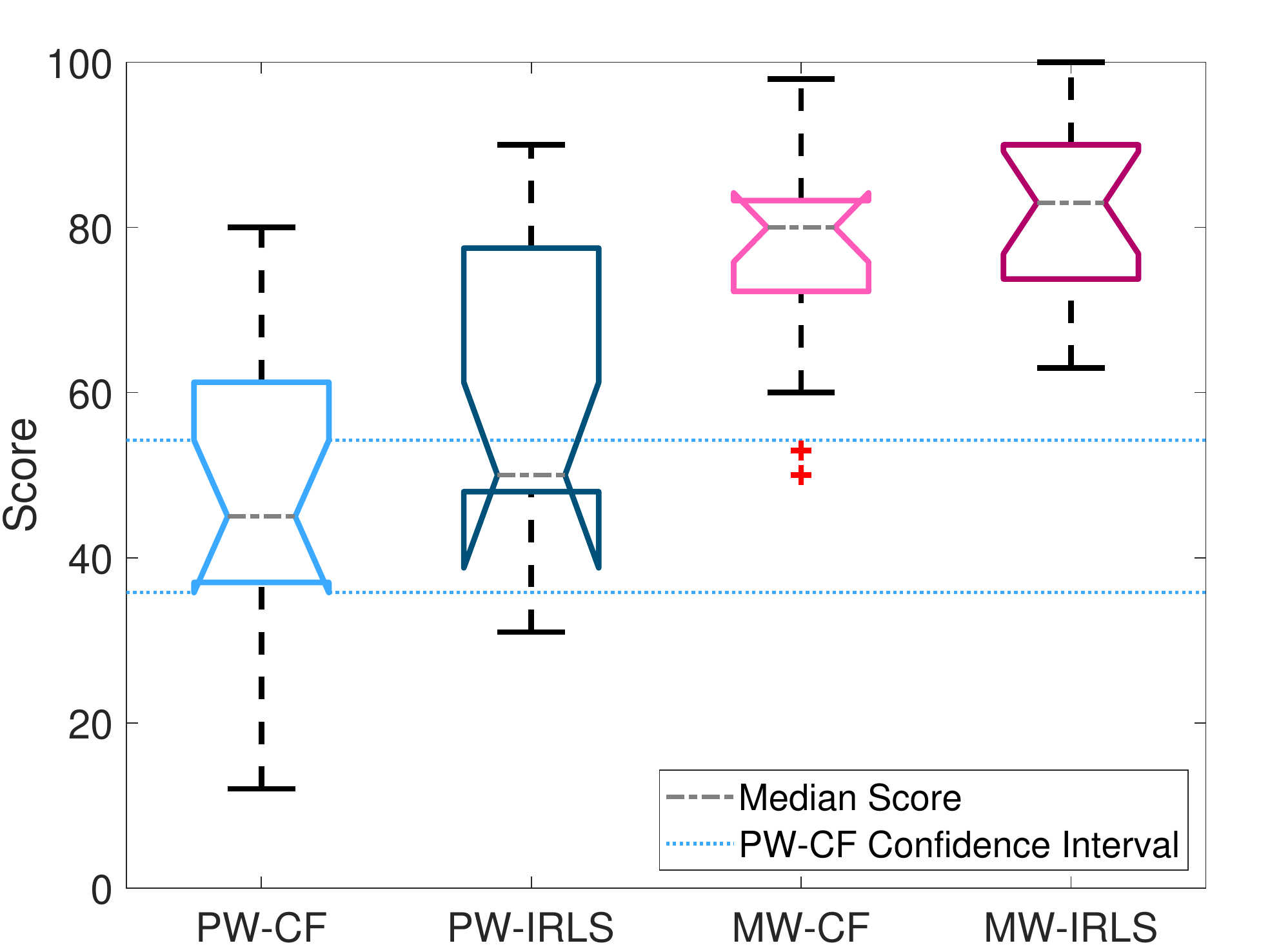}
        \caption{Basic audio quality scores with speech sound-source.}
        \label{fig:spc-baq}
    \end{subfigure}
    \begin{subfigure}{\boxScale\textwidth}
        \centering
        \includegraphics[width=\columnwidth]{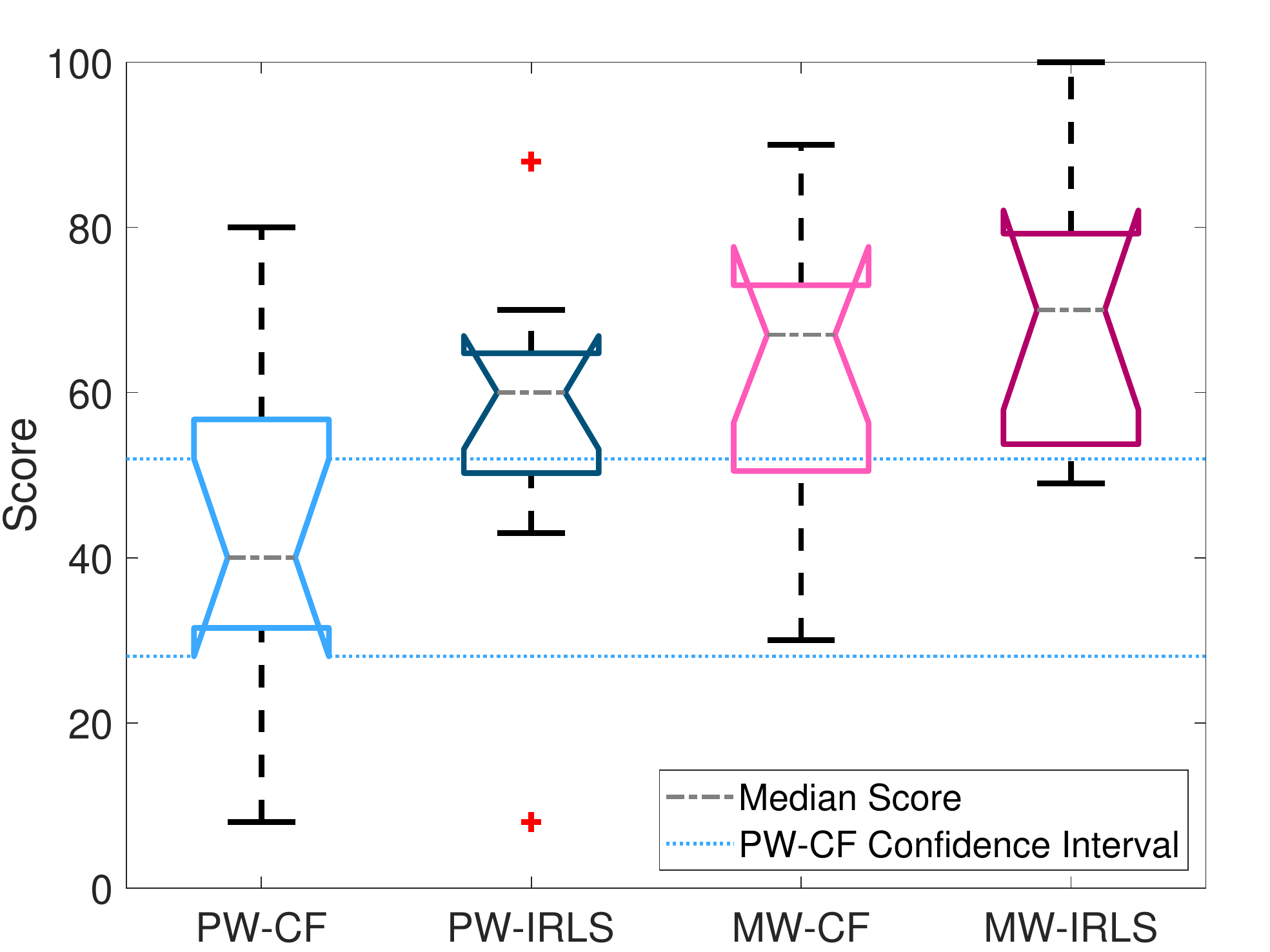}
        \caption{Source localization scores with music sound-source.}
        \label{fig:msc-dir}
    \end{subfigure}
    \begin{subfigure}{\boxScale\textwidth}
        \centering
        \includegraphics[width=\columnwidth]{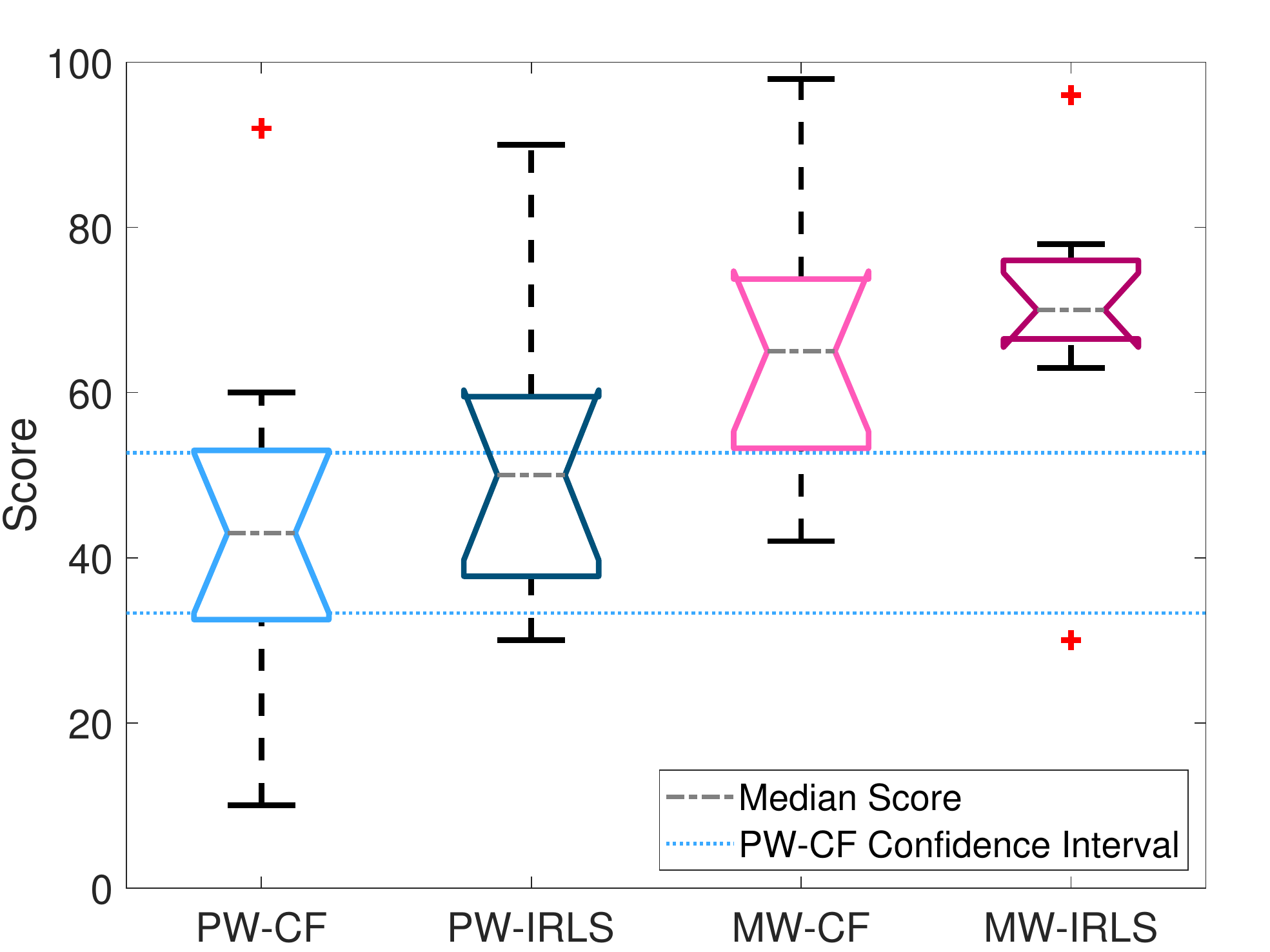}
        \caption{Basic audio quality scores with music sound-source.}
        \label{fig:msc-baq}
    \end{subfigure}
    \caption{Box plot of perception experiment scores for source localization (a) and (c), and basic audio quality (b) and (d). Each box bounds the interquartile range (IQR) with the center bar indicating the median score, and the whiskers extended to a maximum of $1.5\times\text{IQR}$. The \emph{v}-shaped notches in the box refer to the $95\%$ confidence interval. When the notches between two boxes do not overlap, it can be concluded with $95\%$ confidence that the true medians differ.}
    \label{fig:boxplots} 
\end{figure*}

\subsection{Experiment Results}

\subsubsection{Box Plot} 

Figure \ref{fig:boxplots} shows the perceptual scores of the translation methods across all four tests. A Lilliefors test $(p_\text{val} > 0.01)$ showed that our collected scores met the requirement for normal distribution, and a Tukey-Kramer multiple comparison test with $95\%$ confidence was used to determine statistical significance. We discuss the results of these scores through an analysis of variance (ANOVA) examination.

\subsubsection{One-factor ANOVA results} 

We used a one-factor ANOVA to determine if any of the translation methods performed significantly different in each of the perception tests. For speech localization (Fig. \ref{fig:spc-dir}), both MW-CF and MW-IRLS showed a significant improvement in score $(F_{(3,64)} = 6.2, p < 0.001)$ compared to the PW-CF benchmark. Similar results $(F_{(3,64)} = 16.25, p < 0.001)$ are shown for speech quality (Fig. \ref{fig:spc-baq}), where the mixedwave methods were found to be significantly different to PW-IRLS in addition to the benchmark. In the music sound-source tests (Fig. \ref{fig:msc-dir} and Fig. \ref{fig:msc-baq}), only MW-IRLS showed significantly improved means over the benchmark, while MW-CF did not. However, MW-CF was still observed to perform well for music localization in Fig. \ref{fig:msc-dir} and music quality in Fig. \ref{fig:msc-baq} as indicated by the significant median scores. 

\subsubsection{Two-factor ANOVA results} 

We performed a two-factor ANOVA to compare the effects of source-type (planewave and mixedwave) and expansion-type (closed-form and IRLS). In all four tests $(p \leq 0.008)$, mixedwave source distributions were found to score higher means than planewave distributions. Whereas, a significant difference in expansion-type was only found in the speech sound-source tests, with IRLS showing better scores. For music localization $(F_{(1,40)} = 3.36, p = 0.074)$ and music quality $(F_{(1,40)} = 1.07, p = 0.307)$, no significant difference was found between closed-form and sparse expansions. Lastly, no interaction effects $(p \geq 0.382)$ between virtual source-type and expansion-type were found.

\subsubsection{Summary and discussion} 

The proposed MW-IRLS method showed an improvement against the PW-CF benchmark in the perceptual criteria of source localization and audio quality for both a speech and music source. Furthermore, MW-CF also received higher mean scores when reconstructing human speech, and higher median scores for music. When comparing virtual expansion-types, the IRLS expansion was seen to have better quality robustness and localizability for a speech source, but not a music source. This may be explained by the IRLS matching the sparseness of the single human's speech, but not the natural sound of music which is normally generated by multiple sound-sources. Nonetheless, this paper focuses on the modeling of secondary virtual sources. No interaction effect between the source model and expansion-type was found. This indicates that the strong perceptual results achieved by mixedwave methods were not dependent on the expansion-type, and are instead an outcome of the near-field and far-field virtual source mixture. In the next section, we conduct a simulation analysis on the sound fields used in this experiment to gain further insight on properties that may have influenced these strong perceptual results.


\section{Simulation Analysis}
\label{sec:SimulationAnalysis}

In this section we simulate the same virtual environments that were used in the perception test (Section \ref{sec:virtualEnvironment}). We examine their pressure and intensity fields to identify factors that may correlate with perceptual performance. 

\subsection{Error Metrics}

We define the pressure error (PE) and intensity magnitude error (IME) between the true and reproduced sound field as
\begin{equation}
    \left(
    \text{PE} = \frac{|P-\Tilde{P}|^2}{|P|^2}, \hspace{0.3cm}
    \text{IME} = \frac{||\boldsymbol{I} - \Tilde{\boldsymbol{I}}||^2}{||\boldsymbol{I}||^2} 
    \right)
    \times 100(\%).
\end{equation}
where $\boldsymbol{I} = \frac{1}{2} \text{Re}\left(P \boldsymbol{V}^{*}\right)$, and $\boldsymbol{V}^{*}$ is the conjugated sound field velocity.
The intensity direction error (IDE), which is denoted as the acute angle between the true recorded and reproduced intensity fields \cite{shin2016velocity}, is expressed as
\begin{equation}
    \text{IDE} = \arccos\left( \frac{\boldsymbol{I} \cdot \Tilde{\boldsymbol{I}}}{||\boldsymbol{I}|| \cdot ||\Tilde{\boldsymbol{I}}||} \right) / \pi \times 100(\%).
\end{equation}
Additionally, for intensity fields, we also illustrate the true and reproduced intensity unit vector difference $= \boldsymbol{I} / ||\boldsymbol{I}|| - \Tilde{\boldsymbol{I}} / ||\Tilde{\boldsymbol{I}}||$.

\begin{figure}
    \centering
    \includegraphics[width=\columnwidth]{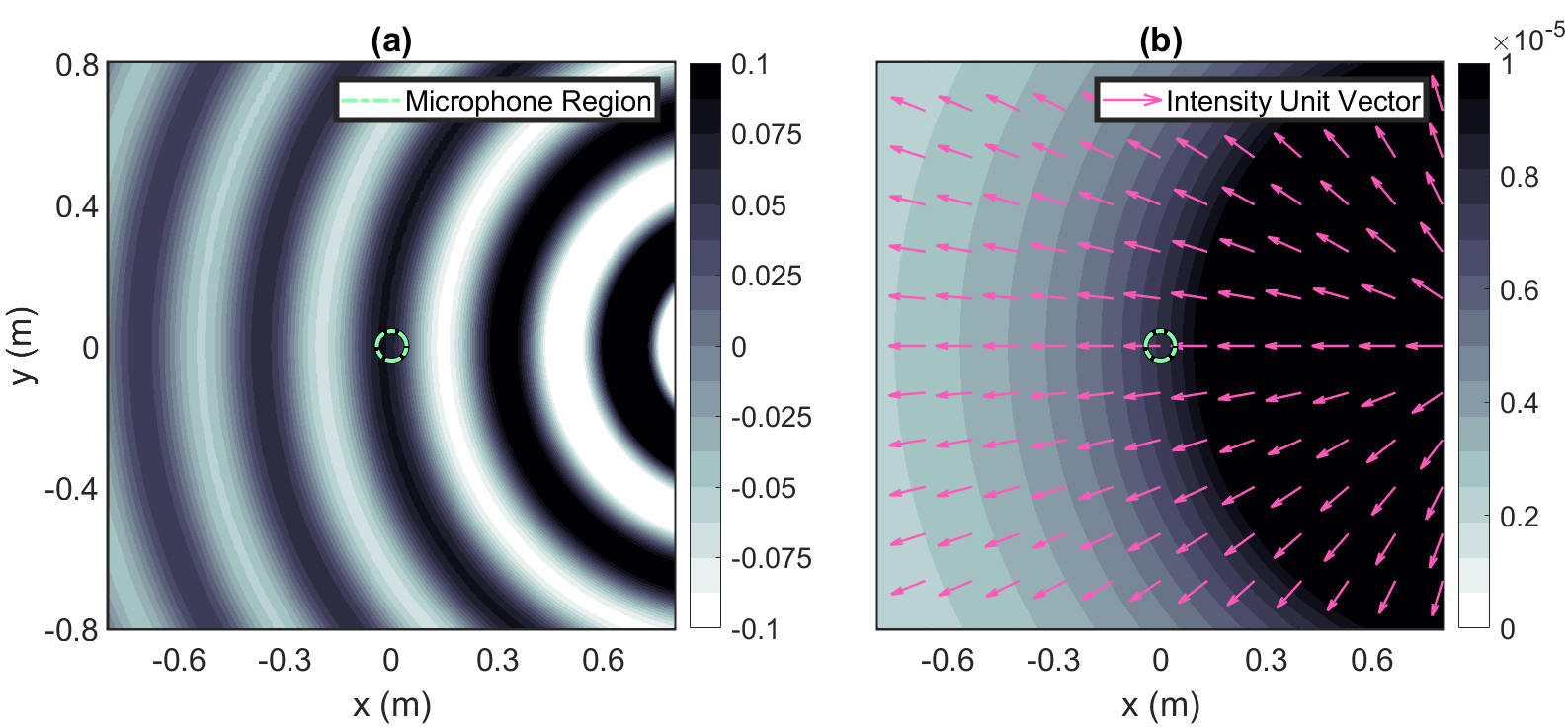}
    \caption{(a) True pressure field and (b) intensity field at $1000\text{ Hz}$ in the XY-plane with the point-source at $\srcPos$, where intensity magnitude is given by the color-map.}
    \label{fig:trueSFD}
\end{figure}

\begin{figure}
    \centering
    \includegraphics[width=\columnwidth]{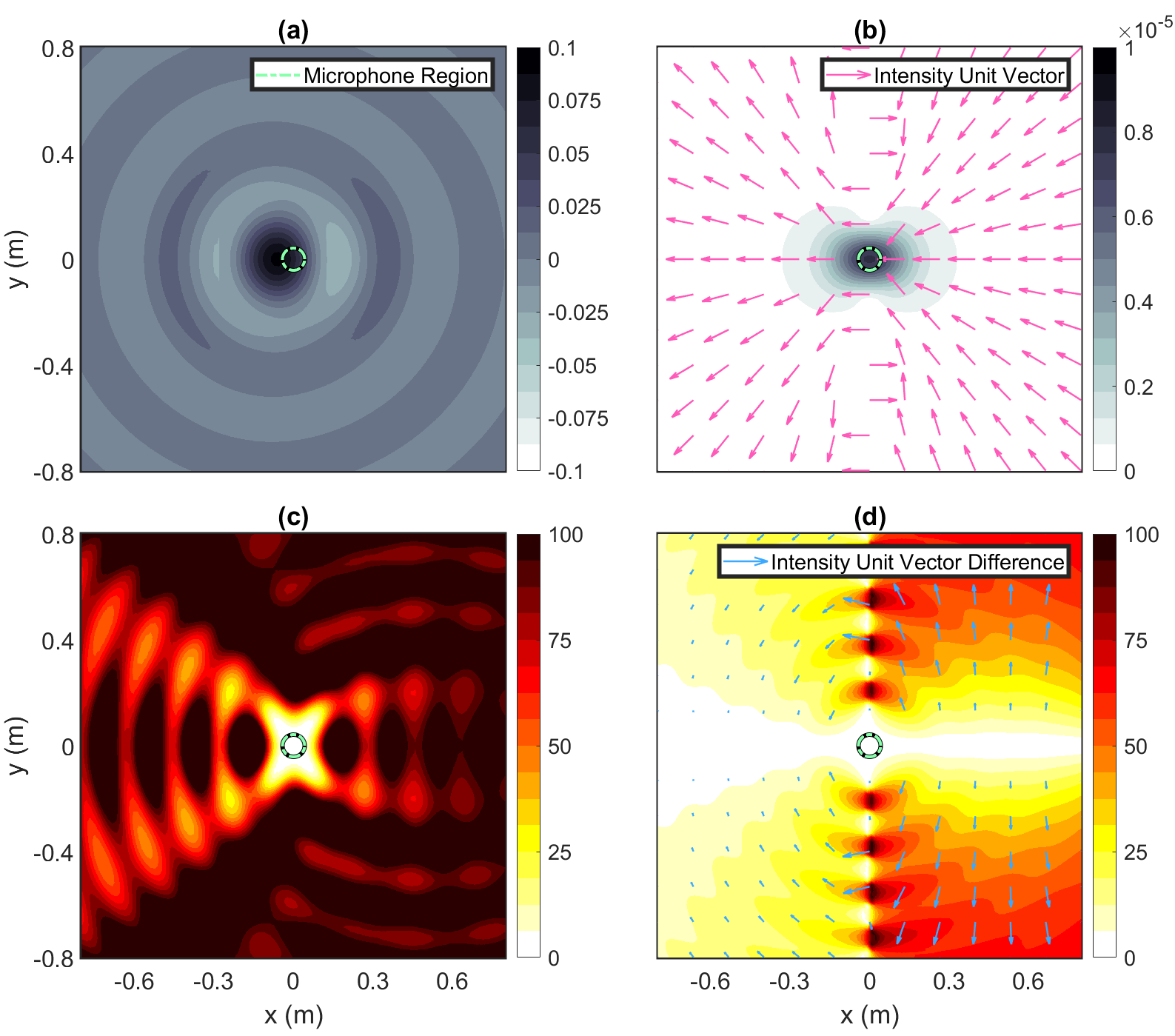}
    \caption{Truncated measurement of (a) the pressure field and (b) the intensity field at $1000\text{ Hz}$ in the XY-plane for the point-source at $\srcPos$, where (c) is PE and (d) is IDE.}
    \label{fig:measSFD}
\end{figure}

\begin{figure}
    \centering
    \includegraphics[width=\columnwidth]{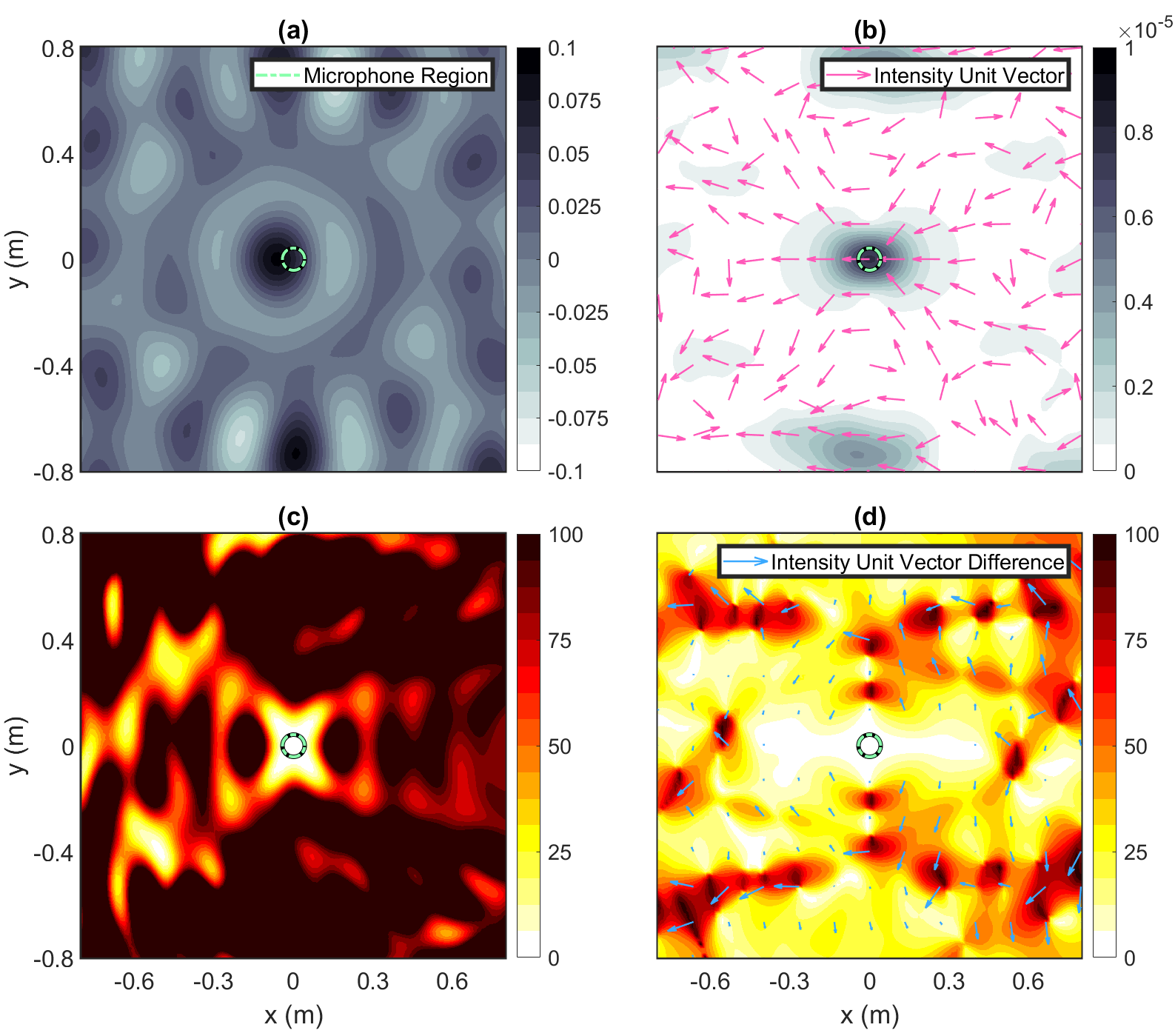}
    \caption{PW-CF reproduction of (a) pressure field and (b) intensity field at $1000\text{ Hz}$ in the XY-plane, where (c) is the reproduction PE and (d) is the reproduction IDE.}
    \label{fig:pwcfSFD}
\end{figure}

\begin{figure}
    \centering
    \includegraphics[width=\columnwidth]{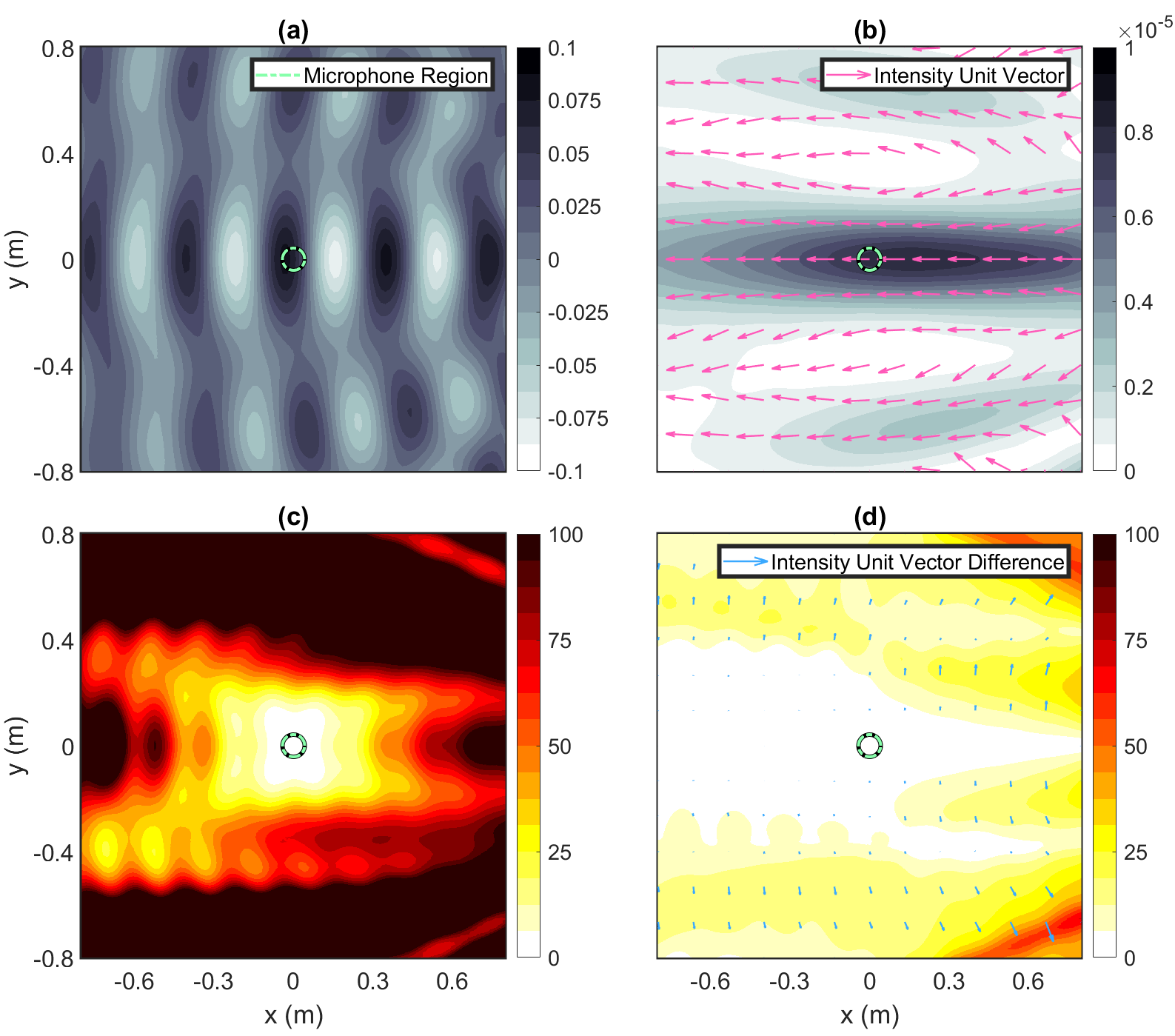}
    \caption{MW-IRLS reproduction of (a) pressure field and (b) intensity field at $1000\text{ Hz}$ in the XY-plane, where (c) is the reproduction PE and (d) is the reproduction IDE.}
    \label{fig:mwirlsSFD}
\end{figure}

\subsection {Pressure and Intensity Fields}

Figure \ref{fig:trueSFD} shows the pressure and intensity field for the true sound-source at $\srcPos$ that we recorded and reproduced virtually in the perception experiment. The $4^\text{th}$ order recording of this true sound-source is shown in Fig. \ref{fig:measSFD}. Immediately we observe the effects of truncation in the recorded pressure field (Fig. \ref{fig:measSFD}a), where a distinct near-field pattern is no longer visible. As expected, the recording is seen to be localized spatially within the microphone array, illustrated by the sweet-spot within the PE (Fig. \ref{fig:measSFD}c). Similarly, the recorded intensity is seen to also be concentrated about the sweet-spot. Beyond the sweet-spot, truncation error is seen to degrade the pressure and intensity accuracy, leading to the perceptual artifacts we wish to resolve by extrapolating a virtual source environment. 

In Fig. \ref{fig:pwcfSFD}, we observe that the PW-CF experiences the same sweet-spot behaviors as the truncated recording, where once again the reproduced pressure and intensity is localized to the microphone's region (Fig. \ref{fig:pwcfSFD}c). A similar result is also obtained by the MW-CF method (not shown), supporting that the sweet-spot is caused by the closed-form expansion over-approximating the truncated recording. The PW-CF intensity field is also seen to be non-uniform throughout the virtual environment. It is expected that this may be a dominant factor contributing to the PW-CF's perceptual evaluation.

Figure \ref{fig:mwirlsSFD} shows better results for the MW-IRLS reproduction. As intended, IRLS expansion is seen to relax the sweet-spot constraint (Fig. \ref{fig:mwirlsSFD}a \& c). Similar results are also observed for the PW-IRLS method (not shown), indicating that sparse expansions are able to extend the region of reproduction accuracy. This is believed to aid the perceptual stability of the reproduction as the listener translates further from the original recording position. Furthermore, the MW-IRLS intensity field (Fig. \ref{fig:mwirlsSFD}b) is shown to have improved uniformity, which leads to better IDE results (Fig. \ref{fig:mwirlsSFD}d). This uniformity is expected to have contributed to the strong perceptual results achieved by the MW-IRLS method.

\begin{figure}
    \centering
    \includegraphics[width=\columnwidth]{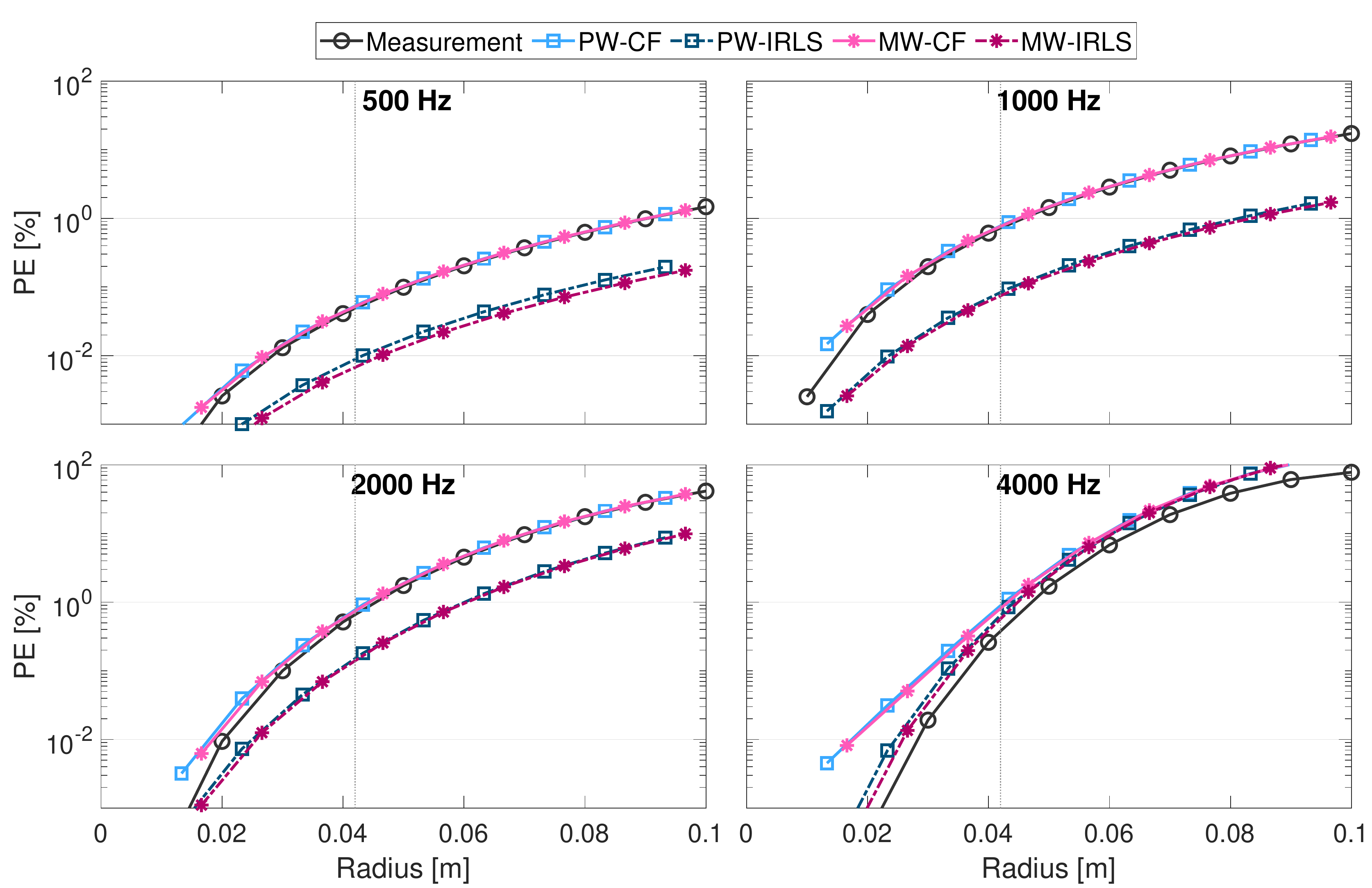}
    \caption{Average pressure error over a spherical surface of varying radius at four frequencies for the measured and reproduced sound fields.}
    \label{fig:avePE}
\end{figure}

\begin{figure}
    \centering
    \includegraphics[width=\columnwidth]{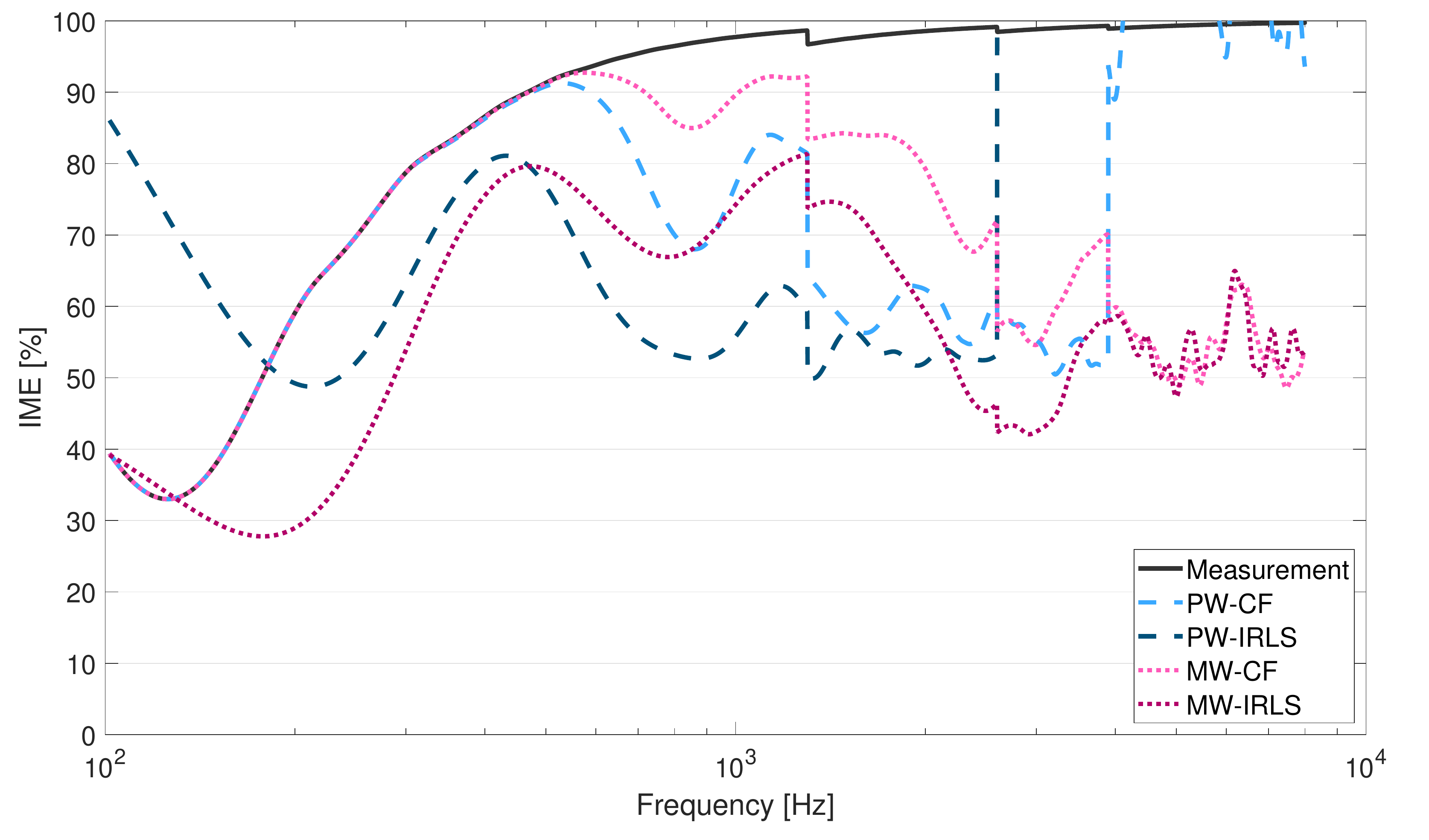}
    \caption{Average intensity magnitude error over a $0.8\text{ m}$ spherical surface plotted against frequency.}
    \label{fig:aveIME}
\end{figure}

\subsection{Pressure and Intensity Error}

We present the averaged PE at various translation distances in Fig. \ref{fig:avePE}. A clear difference in performance is observed at the lower frequencies, where the two IRLS expansions (PW-IRLS and MW-IRLS) are seen to better reproduce the pressure field throughout a $0.1\text{ m}$ region. This result corroborates with the prior sweet-spot observations, where the IRLS expansions are able to relax spatial constraints. On the other hand, the closed-form expansions are shown to match the PE of the recording, further illustrating that the PW-CF and MW-CF methods over-approximate the truncation artifacts.

\begin{figure}  
    \centering
    \includegraphics[width=\columnwidth]{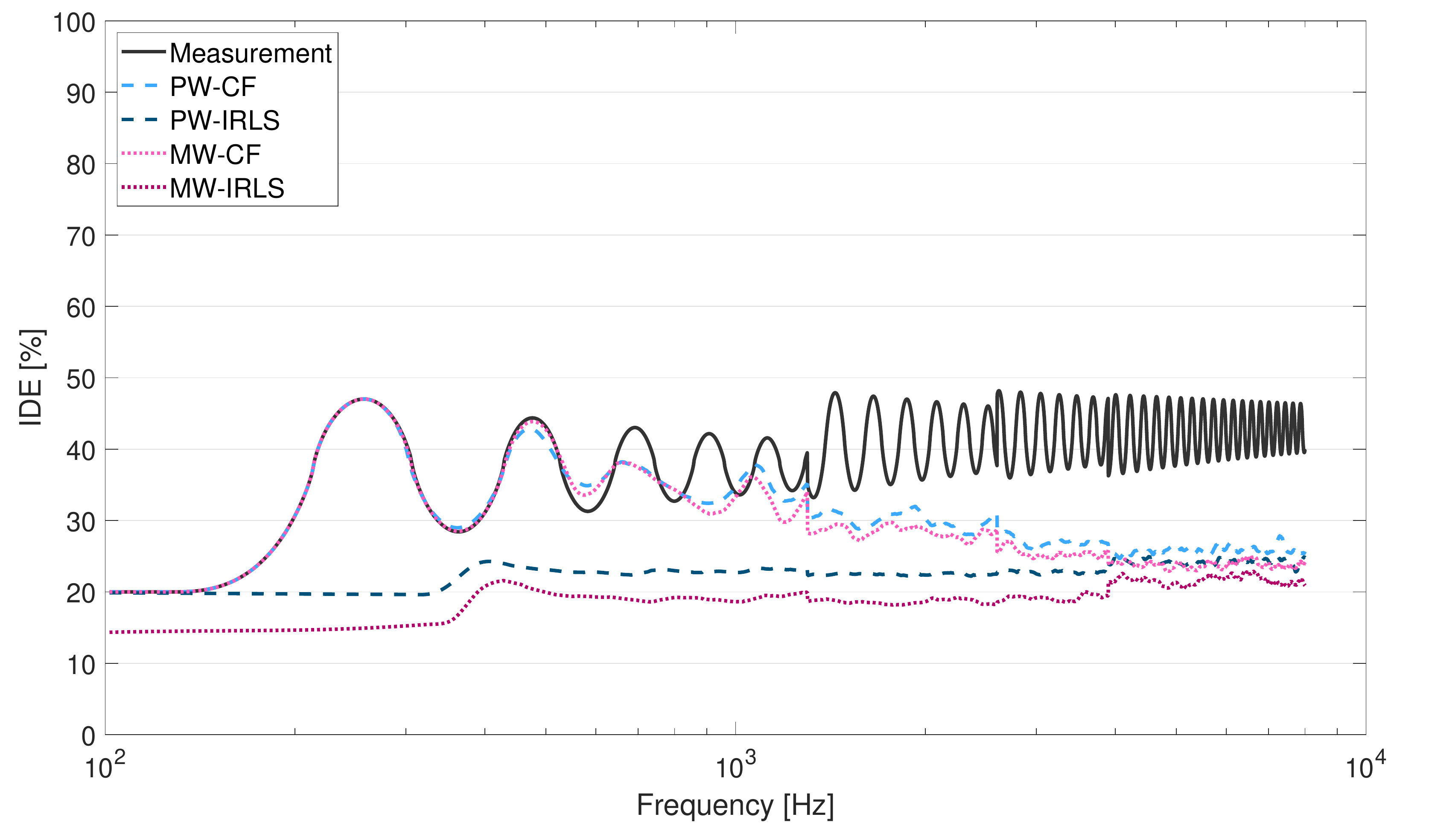}
    \caption{Average intensity direction error over a $0.8\text{ m}$ spherical surface plotted against frequency.}
    \label{fig:aveIDE}
\end{figure}

\begin{figure}
    \centering
    \includegraphics[width=\columnwidth]{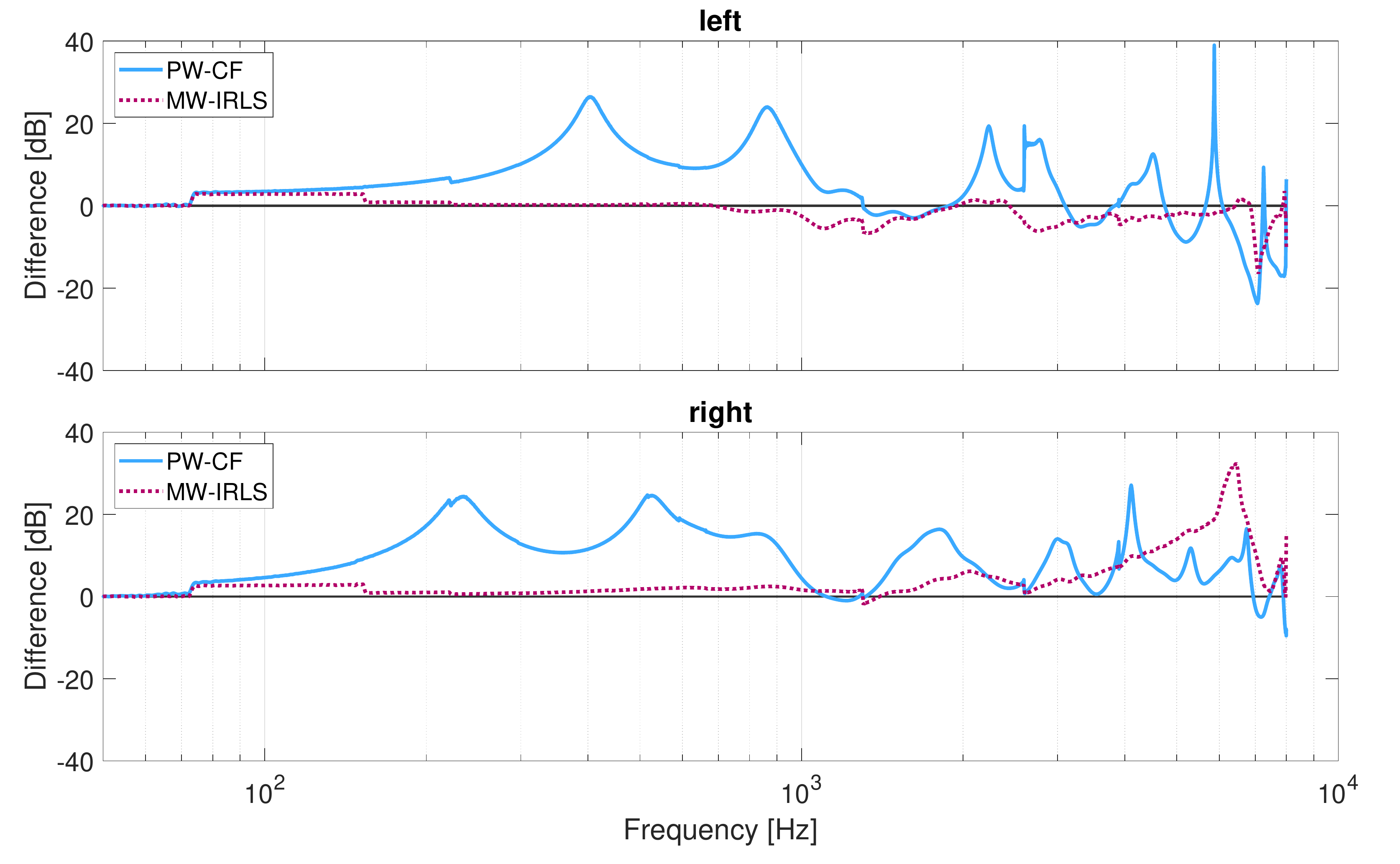}
    \caption{BRIR spectral difference between the true (reference) and reproduced (PW-CF, MW-IRLS) signals rendered at the translated position $(0,0.5,0)\text{ m}$.}
    \label{fig:brirERR}
\end{figure}

All methods are observed to have poor IME at the translation of $0.8\text{ m}$ in Fig. \ref{fig:aveIME}. At higher frequencies both MW-CF and MW-IRLS have lower error than their planewave counterparts. However, the IME is still poor, and it is difficult to know if this behavior contributed to perceptual results. Additionally, large spikes in error are found when the microphone's truncation increases between the $\lceil k |\boldsymbol{x}_Q| \rceil$ frequency bands. It may be possible to smooth the activation of each band to further improve perceptual stability.

The IDE shows clearer results at the $0.8\text{ m}$ translation in Fig. \ref{fig:aveIDE}. The MW-IRLS reproduction is seen to strongly match the direction of the true sound-source's intensity across the full frequency range. This intensity alignment is expected to have contributed to the perceptual results of the MW-IRLS method. This is in contrast to the PW-CF benchmark which is seen to follow the recording's poor IDE at lower frequencies.

\subsection{BRIR Response}

We measure the reproduction system's response by recording, expanding, and auralizing a sine-sweep signal with the planewave and mixedwave translation methods. This provides the binaural room impulse response (BRIR) of the translated listener in the virtual environment. Figure \ref{fig:brirERR} gives the BRIR spectra difference of the PW-CF and MW-IRLS compared to the reference at the translated position of $(0,0.5,0)\text{ m}$ to the left. The BRIR spectral results show the most substantial difference between the PW-CF and MW-IRLS methods thus far. Below $1000\text{ Hz}$ the MW-IRLS is observed to have little spectral deviation from the reference BRIR. This suggests that the MW-IRLS accurately reconstructs the sound heard by a translated listener in the true environment. The PW-CF BRIR, however, is seen to deviate significantly from the reference. Therefore, the BRIR spectra differences support that the MW-IRLS offers greater perceptual accuracy, which was observed in the perceptual experiment results.

The BRIR includes the effects of HRTF processing that is applied to each sound field translation method. This may explain why there is such a large disparity between the planewave and mixedwave BRIR. The mixedwave method adapts the HRTF with the listener's movement. While the planewave method maintains a constant HRTF perspective and shifts the truncated reproduction about the listener. It is the difference between these two HRTF implementations that may have the most significant effect on the BRIR differences and the perceptual experiment results.


\section{Conclusion}
\label{sec:conclusion}

Virtual reality technology enhances acoustic real-world reproductions by allowing listeners to perceptually move about the environment. At this time, however, the benchmark planewave method towards sound field translation is still limited by inherited microphone constraints. Furthermore, the planewave source model is restricted to the far-field, which results in the listener's HRTF perceptive being fixed during translation. As a result, immersion in the planewave environment is degraded by poor source localizability and audible spectral distortions.  

We have proposed an alternative source model for translation that enables a sparse virtual environment to contain a mixture of near-field and far-field sources. We compared this proposed mixedwave method against the planewave benchmark through a perceptual MUSHRA experiment and cross-examined the results with numerical simulations. For human speech reproduction, the mixedwave source model improved both localizability and audio quality. Both the closed-form and IRLS expanded mixedwave reproductions were found to provide a more immersive experience. Similar results were also found for a wider band music sound source. 

The IRLS expansion was shown to help enlarge the reproduction sweet-spot, and in response it scored better perceptually than the closed-form expansion for speech sound. Additionally, the proposed method showed better intensity direction matching than the benchmark, further corroborating the perceptual results. Finally, we illustrated that the mixedwave's ambisonic-like binaural rendering allows for greater perceptual accuracy due to lower BRIR spectral error. 

We note that this paper focuses on the sole effects of modeling a near-field far-field mixture for translation. As such, we studied an over-simplified acoustic environment in order to make clearer comparisons. We leave the considerations involved with implementing the proposed method as future work. Accounting for acoustic reflections, diffuse sound, multiple sound sources, source directivity, and the methods to separate and process them in a virtual mixedwave environment are left as an open problem. Furthermore, it may be satisfying to reproduce the recorded experience along with synthetic sounds, and it should be explored along side future applications developed for virtual reality. 


\section{Thanks}
The authors would like to thank Zamir Ben-Hur for guidance in developing the perceptual test, and Shawn Featherly for development of the perceptual test Unity application.

\bibliographystyle{IEEEbib}
\bibliography{main}

\end{document}